\documentclass[review,12pt]{elsarticle}
\usepackage{amssymb}
\usepackage{amsfonts}
\usepackage{amsmath}
\usepackage{color}
\usepackage{graphicx}
\usepackage{natbib}
\usepackage{caption}
\usepackage{float}

\journal{Journal of Non-Newtonian Fluid Mechanics}
\begin{document}
\begin{frontmatter}

\title
{The folding motion of an axisymmetric jet of wormlike-micelles solution}

\author{Matthieu Varagnat~$^{a}$, Trushant Majmudar~$^{a}$, Will Hartt~$^{b}$, Gareth H. McKinley~$^{a}$}

\address{$^a$ Department of Mechanical Engineering,
Massachusetts Institute of Technology, Cambridge, 02139 MA, USA}
\address{$^b$ Corporate Engineering Technologies Lab, The Procter and Gamble Co, West Chester, OH 45069}

\begin{abstract}

The problem of buckling and coiling of jets of viscous, Newtonian liquids has received a substantial level of attention over the past
two decades, both from experimental and theoretical points of view.
Nevertheless, many industrial fluids and consumer products are non-Newtonian, and
their rheological properties affect their flow behavior. The
present work aims at studying the dynamics of cylindrical jets of a viscoelastic, 
shear-thinning solution of cetylpyridinium salt
(CPyCl). In concentrated solutions, CPyCl surfactant molecules have been shown
to assemble in long wormlike micellar structures, which gives the fluid its
non-Newtonian properties. Jets of this fluid show novel features compared to
their Newtonian counterparts, including a type of motion, in which the jet
folds back and forth on itself in a fashion similar to sheets of viscous fluids, instead of coiling
around the vertical axis as cylindrical Newtonian jets do. Another novel
feature of CPyCl micellar fluid jets is a widening of the jet above the plate
reminiscent of the die-swell phenomenon that we call \emph{reverse swell}. We
propose physical mechanisms for both folding and reverse swell, and compare
theoretical predictions to experimental measurements. In addition, we systematically
explore different flow regimes in the parameter space of the height of fall and flow
rate and compare regime maps of a CPyCl micellar solution and a Newtonian
silicone oil.
\end{abstract}
%


\end{frontmatter}%

\section{Introduction}

Situations where a continuous stream of material is flowing downward at
a moderate velocity onto a plane surface occur in both everyday life and
industrial applications. Honey poured on a toast, the filling of a shampoo
bottle in an automated line, as well as numerous other examples in the oil,
food, and cosmetic industries, are similar problems in which the stability of
the streaming jet is a crucial feature. From an industrial point of view, stable jet spreading homogeneously on the support, is almost
always preferred. Indeed, jet instabilities can lead to problems such as the
entrapment of air bubbles in the folds of a buckled planar jet as described by
Pouligny \cite{2008Pouligny} or the stacking and mounding of yielding materials such as pastes
or emulsions. There is therefore a strong practical motivation for understanding the emergence of different flow regimes in this situation and the relevant control parameters.

\begin{figure}
[!ht]
\begin{center}
\includegraphics[width=450pt]
{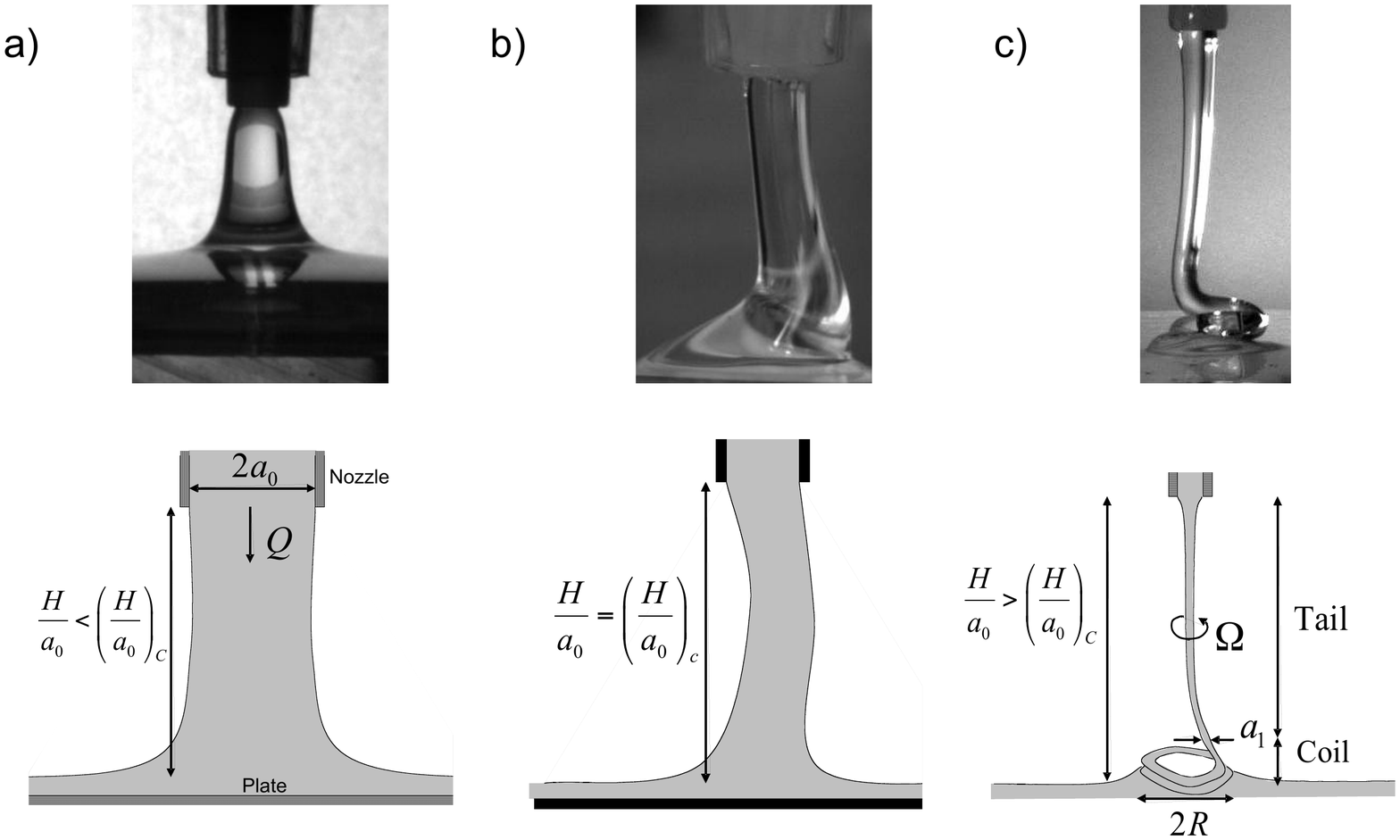}
\caption{}
\label{schemes}
\end{center}
\end{figure}

Depending on the fluid properties, and the control parameters such as the flow rate
$Q$ and the height of fall $H$, a jet can exhibit different behaviors or
regimes. At low heights of fall, the jet spreads homogeneously on the plate forming 
a steady stagnation flow, as
shown in Fig.~\ref{schemes}(a). Cruickshank and
Munson \cite{1982CandMsteady} have provided an analytical solution for the shape of
the jet, driven by a balance between the imposed flow rate, gravitational
acceleration, and the viscous resistance to deformation of the fluid. 
The shape of the jet is controlled by the parameter $\beta=H a_{0} \sqrt{g/6Q\nu}$,
where $a_{0\text{ }}$is the nozzle radius, $\nu$ is the kinematic viscosity
and $g$ is the acceleration of gravity. For $\beta>\pi/2$, the jet starts to
thin under the influence of gravity instead of monotonically enlarging while
spreading on the plate. For heights of fall large enough so that $\beta\gg
\pi/2$, the jet radius reaches a limit $a_{1}$ for which the acceleration of
gravity is balanced by the viscous resistance, and the balance of the two
forces leads to
\begin{equation}
a_{1}\sim\sqrt{\frac{3\nu Q}{Hgl_{VG}}}\label{a1Dimensional}%
\end{equation}

where $l_{VG}=(\nu^{2}/g)^{1/3}$ is the characteristic length scale over which
gravitational and viscous effects balance each other.

As the height of fall is increased, the jet leaves the stagnation flow regime and 
starts buckling under the compressive viscous stress, as shown in Fig.~\ref{schemes}(b). 
This situation was first described by
Taylor \cite{1968Taylor} and was studied experimentally by Cruickshank and
Munson \cite{1981CandM} for planar and axisymmetric jets.
The theoretical framework developed by Cruickshank \cite{1988Cruickshank} and
Tchadarov \cite{1993Tchavdarov} for the axisymmetric geometry and by Yarin and
Entov \cite{1995Yarin} for the planar case is in good agreement with these experimental results. Cruickshank showed \cite{1988Cruickshank} that the axisymmetric jet can buckle according to
two modes, azimuthal and non-azimuthal (called coiling and
folding respectively, in this article). They derived a critical buckling height for each
mode, given by the following geometrical conditions%
\begin{align}
\frac{H}{2a_{0}} &  =7.663\text{ (coiling)}\label{coilbcklimit}\\
\frac{H}{2a_{0}} &  =4.810\text{ (folding)}\label{foldbckllimit}%
\end{align}

The validity of Eqs.~(\ref{coilbcklimit}) and (\ref{foldbckllimit}) have been confirmed experimentally by Cruickshank and Munson \cite{1981CandM}. They showed that a cylindrical jet of Newtonian fluid transitions from a steady jet to a coiling one (Fig.~\ref{schemes}(c)) at a height given by Eq.~(\ref{coilbcklimit}), and that a planar jet starts folding when the condition in Eq.~(\ref{foldbckllimit}) is satisfied. However, it was also observed that these transitions are suppressed at higher Reynolds numbers \citep{1981CandM, 1988Cruickshank}. Beyond a critical Reynolds number, an axisymmetric or planar jet exhibits stagnation flow for all nozzle-to-plate distances. The Reynolds number is defined here as the ratio of viscous timescale $t_{V}=a_{0}^{2}/\nu$ to convective timescale  $t_{I}=\pi a_{0}^{3}/Q$:

\begin{equation}
Re=\frac{t_{V}}{t_{I}}=\frac{Q}{\pi\nu a_{0}}\label{ReTimescales}%
\end{equation}

For axisymmetric jets, the critical Reynolds number above which coiling of the jet disappears is given by:

\begin{equation}
Re_{crit}=\frac{Q}{\pi\nu a_{0}}~=~1.2
\end{equation}

For planar jets, the critical Reynolds number above which folding of the jet disappears is given by:

\begin{equation}
Re_{crit}=\frac{Q}{\pi\nu a_{{\rm s}}}~=~0.56
\end{equation}
where, $a_{{\rm s}}$ is the slit radius.

Furthermore, in the cylindrical case, they reported that below this limit, for  $0.1<Re<1$, two transitions were observed as the height of fall was progressively increased; from a stable jet to folding, and from folding to coiling. The first buckling transition happens roughly at the critical height predicted by Eq.~\ref{foldbckllimit}, and the height of the folding-coiling transition is approximately given by Eq.~\ref{coilbcklimit}. These subtle transitions are difficult to observe experimentally for two reasons; the first reason is that the flow rates required to reach this range of Reynolds number are high, and the second reason is that the values of the scaled height $H/2a_{0}$, where these transitions occur, are very low. Both of these factors create a situation where a jet whose diameter is almost the same as the nozzle diameter, oscillates under high flow rates at very low nozzle-to-plate distances, making it very difficult to distinguish between coiling and folding.

\begin{figure}
[!ht]
\begin{center}
\includegraphics[
width=400pt
]
{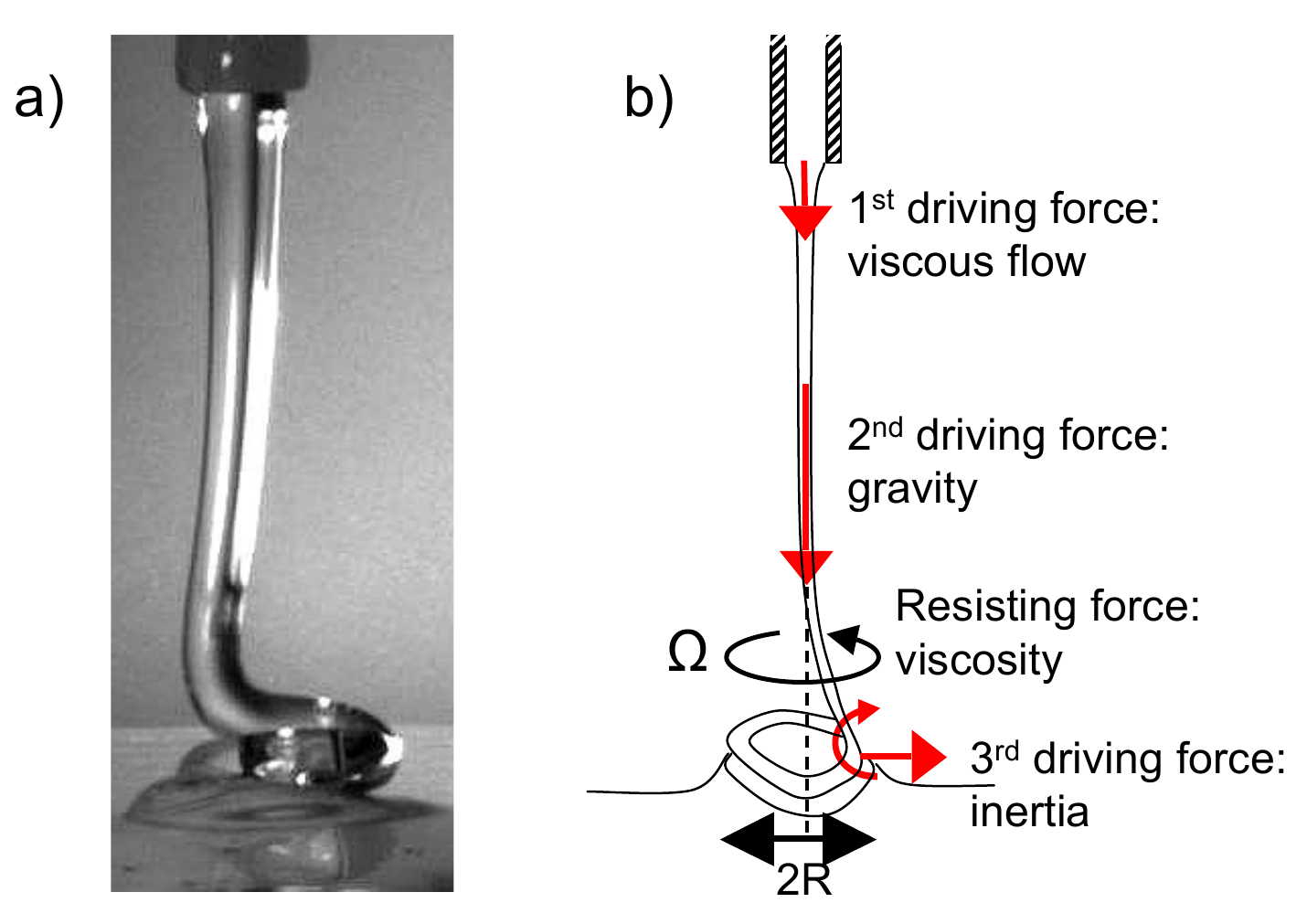}
\caption{}
\label{allGraphsMechanism}
\end{center}
\end{figure}

After the onset of buckling, the jet of a viscous Newtonian
fluid starts to coil around the vertical axis, as can be seen in Fig.~\ref{allGraphsMechanism}(a). 
In the limit of large heights of fall, Mahadevan
and co-workers \cite{1998MahaVJ, 2000MahaCorrection} derived 
expressions for the coiling radius and frequency by balancing 
viscous and inertial forces in the coil; the viscous forces arise from the
curvature and bending of the jet in that region. Ribe \cite{2004RibeVJ} showed that this
analysis was a subset of a broader picture, with three distinct regimes,
viscous, gravitational and inertial, depending on which forces were dominant.
Ribe's simulations for the frequency of
coiling as a function of height, at a given flow rate, were in good agreement with experiments by Maleki and coworkers 
and showed multi-valued frequencies at the transition between gravitational
and inertial regimes \cite{2004RibeVJmultival}.

Under the assumption that the jet radius is constant in the coil, the frequency,
coiling radius, and the final radius of the jet, are connected by the conservation
of volume
\begin{equation}
Q=\pi\Omega Ra_{1}^{2}\label{volumeConservation}%
\end{equation}

Figure~\ref{allGraphsMechanism}(b) schematically details all the relevant forces along the jet length. 
The resisting torque is always due to viscosity. Viscous bending stresses arise
because of the velocity gradient between the inner part of the curved jet, 
where the velocity is smaller, and the outer part where it is maximum. The
velocity gradient scales like $U_{1}a_{1}/R^{2}$, where $U_{1}=Q/\pi a_{1}
^{2}$\ is the axial velocity just before the coil. In a fashion similar to
beam-bending in solid mechanics, the viscous stress $\eta Q/\pi R^{2}a_{1}$
integrated over the jet cross-section vanishes, but the integrated torque
remains non-zero and scales as \cite{2000MahaCorrection}
\begin{equation}
T_{coil}\sim\frac{\eta Qa_{1}^{2}}{R^{2}}\label{viscousTorque}%
\end{equation}

\subsection{ Viscous regime}

At low heights of fall, the whole tail of the jet is bent sideways and the viscous torque caused by the fixed
vertical orientation of the nozzle controls the motion of the jet (the first driving force in Fig.~\ref{allGraphsMechanism}(b)). 
This is also a shear-induced
torque, with a characteristic curvature $1/H$, that scales like $\eta
Qa_{1}^{2}/H^{2}$. The torque balance leads to the scaling
laws for the coiling radius $R_{V}$ and the coiling frequency $\Omega_{V}$ \cite{2004RibeVJ} given by:
\begin{subequations}
\begin{align}
R_{V} &  \sim H\label{ViscousRad}\\
\Omega_{V} &  \sim\frac{Q}{a_{0}^{2}H}\label{ViscousFreq}%
\end{align}

In this regime, the jet motion is controlled by external parameters such as
the height of fall and imposed flow rate, whereas in the other two regimes the
fluid jet selects its own dynamics through a balance of forces involving intrinsic
fluid properties as well as external parameters.
\end{subequations}
\subsection{ Gravitational regime}

At larger heights of fall, only the lowest part of the tail is bent, on a
length scale of the radius of the coil ($R$). The weight of the fluid in this part
is of the order of $\rho gRa_{1}^{2}$, the lever arm is of the order $R$, and hence the
buckling torque is $\rho gR^{2}a_{1}^{2}$. Using the scaling in Eq.~
(\ref{a1Dimensional}) for $a_{1}$, the torque balance gives 
the following scaling laws for the coiling radius $R_{G}$ and the frequency $\Omega_{G}$ \cite{2004RibeVJ}:
\begin{subequations}
\begin{align}
R_{G} &  \sim\left(  \frac{\nu Q}{g}\right)  ^{\frac{1}{4}}\label{GravRad}\\
\Omega_{G} &  \sim H^{2}\left(  \frac{g^{5}}{\nu^{5}Q}\right)  ^{\frac{1}{4}%
}\label{GravFr}%
\end{align}

The transition from viscous to gravitational regime happens \cite{2004RibeVJ}
for $\Omega_{G}\simeq2\Omega_{V}$, or $H_{VG}^{{}}\simeq(Q\nu/g)^{5/12} 
a_{0}^{-2/3}$. Note that if this height $H_{{\rm VG}}^{{}}$ is lower than the
buckling height $H_{{\rm buckling}}=7.663\times2a_{0}$, then coiling simply starts
in gravitational mode. Indeed, for typical experimental values of ~$\nu
=200$~cm$^{2}/s$, $a_{0}^{{}}=1.25$ mm, and $Q=3$~mL/min, $H_{\rm VG}\simeq0.5$ cm,
while $H_{\rm Buckling}\simeq1.9$ cm. Note also that a simpler force balance
between gravitational and viscous forces given by
$\beta\approx1$, leads to the qualitatively similar result of $H\simeq
(Q\nu/g)^{1/2}a_{0}^{-1}$.
\end{subequations}
\subsection{Inertial regime}

When the rotational inertia in the coil becomes important, it can drive the
coiling as well. The inertial force in a rotational reference frame scales per
unit volume as $\rho\Omega^{2}R$, so the torque scales as $\rho\Omega^{2} 
R^{3}a_{1}^{2}$. The torque balance together with Eqs.~(\ref{a1Dimensional}) and
(\ref{volumeConservation}) lead to the following scaling laws
for the coiling radius $R_{I}$ and frequency $\Omega_{I}$ \cite{2000MahaCorrection}:
\begin{subequations}
\begin{align}
R_{I} &  \sim\nu g\left(  \frac{Q}{H^{4}}\right)  ^{\frac{1}{3}}%
\label{InerRad}\\
\Omega_{I} &  \sim\frac{1}{\nu^{2}}\left(  \frac{g^{5}H^{10}}{Q}\right)
^{\frac{1}{3}}\label{InerFr}%
\end{align}

This regime becomes the most stable one when $\Omega_{I}>2\Omega_{G}$ 
\cite{2004RibeVJ}, i.e. for $H_{\rm GI}\simeq\left(  Q\nu^{9}/g^{5}\right)
^{1/16}$. For the values used above, this represents a height of approximately $H=11.0$ cm. Note that
for $\Omega_{G}<\Omega_{I}<2\Omega_{G}$, the system is multivalued, with the
solution oscillating between the two possible frequencies.
In all of the above analyses the relevant surface tension parameter comparing the
capillary thinning process to gravity is much smaller than unity 
$\sigma/\rho ga_{0}^{2} << 1$, where $\sigma$
is the surface tension of the fluid. Cruickshank
and Munson have reported experimental data for a surface tension parameter
slightly larger than one \cite{1981CandM}, but never to a point where the 
destabilizing action of capillary forces breaks the jet into droplets.

So far, the research on jets impacting a plate has mostly focused on Newtonian
fluids, for example with viscous fluids such as silicone oil. Nevertheless,
real-life fluids are almost always more complex: pastes, gels, and surfactants are ubiquitous in
the healthcare, cosmetic, and food industries. Rheologically complex fluids are used in many aspects of everyday life; from ketchup
and mayonnaise to foaming detergents, skin creams, and hair conditioners.
Beyond the industrial interest toward extending the field of study to
non-Newtonian fluids, the motivation for this work is to connect the
rheology of the fluid to its jetting properties. We focus on jets of 
wormlike micellar solutions of cetylpyridinium chloride (CPyCl). 
Wormlike micellar fluids are non-Newtonian fluids that have recently attracted
a lot of attention for their strong, tailorable viscoelastic and
shear-thinning properties, their ease of use, and the possibility they offer
as model fluids for surfactant-based consumer products \cite{1988Rehage}.

Wormlike micellar fluids are concentrated aqueous solutions of one or more
ionic surfactants, as well as a counter-ion salt. At very low surfactant
concentration, the solution is homogeneous, but when the concentration
increases above a critical threshold, denoted the critical micellar concentration or
CMC, the surfactant molecules spontaneously assemble in spherical structures.
In these spherical micelles the surfactants orient their
organic tails towards the center and bear their ionic heads outside, satisfying
both tail segregation and ionic repulsion. Adding a large
enough counter-ion reduces the curvature of the optimal shape, by screening
the ionic interactions. The change of optimal curvature leads, for some
combinations of surfactants and counter-ions, to changes in the shape of the
assembly \cite{1976Israelachvili}.  In some parts of their phase diagrams \cite{2008Rothstein}, some carefully chosen systems exhibit long linear structures, called wormlike
micelles. The formation of wormlike micellar structure is schematically represented in Fig.~3. 
In studies of extensional flow of these fluids, Yesilata and co-workers \cite{2006McKinley} have used  solutions
of erucyl bis(2-hydroxyethyl) methyl ammonium chloride (EHAC), and Rothstein \cite{2003Rothstein} used
cetyltrimethylammonium bromide (CTAB). Rehage and Hoffmann \cite{1988Rehage}, as
well as Berret and co-workers \cite{CPyCl} have used a solution of CetylPyridinium
Chloride (CPyCl) and a salicylic salt (NaSal) in brine and we employ this well characterized system in the present study.

Wormlike structures at high enough concentrations in surfactant and salt are
long enough to entangle, therefore giving pronounced viscoelastic properties
to the solution. Stress relaxation with wormlike micelles can operate via 
two processes: the first one is reptation, similar to the stress relaxation
process of polymers described by de Gennes \cite{DeGennes}, and the second one
is a breakup-recombination process that is unique to these polymer systems. The characteristic timescales for stress relaxation and breakup-recombination processes are
$\lambda_{r}$ and $\lambda_{b}$, respectively, and Cates showed that, for $\lambda_{b}%
\ll\lambda_{r}$, the linear viscoelastic properties are well described by a
single-relaxation-time Maxwell model with a characteristic timescale $\lambda$ given by \cite{Cates}:
\end{subequations}
\begin{equation}
\lambda=(\lambda_{r}\lambda_{b})^{\frac{1}{2}}\label{LambdaMicelles}%
\end{equation}

In steady shear flow, wormlike micellar solutions are strongly nonlinear fluids,
with a characteristic shear rate $1/\lambda$ and a zero-shear-rate viscosity
$\eta_{0}$ that is strongly dependent on the salt concentration. For shear
rates above $1/\lambda,$ it is found that the shear stress exhibits a plateau and 
remains almost constant at
a critical stress of the order of $\lambda\eta_{0}$. In addition, both
$\lambda$ and $\eta_{0}$\ are observed to be strongly temperature dependent,
with an Arrhenius-type dependence with a very large activation energy. The
strong variations of rheological properties with temperature come from two
effects acting in the same direction. The first is a polymer-like behavior in
which reptation, a thermally activated process, occurs more rapidly at higher
temperature. The second is specific to supramolecular structures: the breakup
rate increases with temperature, which is not only a stress relaxation process
on its own, but also tends to shorten the average length of wormlike micelles,
thus resulting in  a shorter path-length for reptation to occur.

Extensional rheological studies of wormlike micellar solutions conducted using capillary breakup experiments (CaBER) (\cite{2006McKinley},
\cite{CABERCPyCl}) have shown a significant extension strengthening of the
fluid, likely due to the alignment of the wormlike micelles in the extensional 
flow \cite{2003Rothstein}, quite like what happens for dilute polymer
solutions. For example, in the case of CPyCl micellar solutions, Bhardwaj and
co-workers have shown \cite{CABERCPyCl} that the fluid first undergoes an
initial phase of Newtonian visco-capillary thinning \cite{2006McKinley}. As the
liquid bridge thins, the extension rate ($\dot{\epsilon}$) increases,
until the local Weissenberg number $Wi_{{\rm mid}}=\lambda\dot{\epsilon}=0.5$. At
that point the wormlike micelles are stretched too rapidly to
relax, which leads to extension thickening, with elastic stress buildup
resisting the breakup of the thin filament. In contrast to polymer solutions
however, when the local tensile stresses become too large for the micelles to
resist,  \emph{en masse} rupture of the entangled chains results in filament
breakup \cite{2003Rothstein}. In addition, thin liquid bridges and jets have a
large area-to-surface ratio, and therefore solvent evaporation can tend to
cool the fluid and increase the surfactant concentration, both effects
increase the viscosity and lower the thinning rate.

Jets of CPyCl wormlike micellar solution falling on a plate show two novel
features. The shape of the jet is different from its Newtonian counterpart,
with a widening of the jet immediately above the plate, as we have shown in Fig.~\ref{all_scheme_cpycl}(a). This feature is reminiscent of the well
studied die-swell phenomenon \cite{Tanner}, in which jets of non-Newtonian
fluids swell at the exit of a nozzle, and we refer to it as \emph{reverse
swell}. The first goal of this paper is to understand this novel phenomenon
and to predict the amount of swell measured experimentally.
The second feature observed in the viscoelastic CPyCl micellar jets is a 
novel type of dynamic behavior, in which the buckled cylindrical jet remains 
in a vertical plane and folds back and forth on itself, as shown 
in Fig.~\ref{all_scheme_cpycl}(b). For
Newtonian fluids the folding motion is observed only for planar jets and never
for cylindrical jets. This new type of motion dominates in most experimental
cases, but for some values of the experimental parameters, jets of CPyCl
micellar solutions can coil instead (see Fig.~\ref{all_scheme_cpycl}(c)). The
second goal of this paper is to map these different types of instabilities onto
a suitable non-dimensional parameter space, and describe a mechanism for this viscoelastic folding. We then derive scaling laws for the amplitude and frequency of folding based on the proposed mechanism, and compare them to the experimental data.

This paper is divided in two sections, theoretical and experimental. We first
begin by defining a suitable non-dimensionalization scheme, that allows us to compare
CPyCl solutions at different concentrations and contrast the response with other fluids. We
then propose a mechanism for the reverse swell phenomenon and the folding behavior of
CPyCl jets. We also derive appropriate scaling laws for the amplitudes and frequencies of folding states. Then, we describe the details of the experiments, 
and compare theoretical predictions to experimental data.
\section{Theoretical analysis}

\subsection{Dimensional analysis}

The forces in balance along the flow in the jet are (i) gravity that tends to
accelerate and stretch the jet, and (ii) viscosity that resists the process,
provided that the height of fall is large enough so that $\beta\gg\pi/2$. This
competition can be characterized by a time scale $t_{\rm VG}=(\nu_{0}
/g^{2})^{\frac{1}{3}}$ and a length scale $l_{\rm VG}=(\nu_{0}^{2}/g)^{\frac{1}
{3}}$, where $\nu_{0}$ is the zero-shear-rate kinematic viscosity. These
scalings will typically be relevant as long as strain-hardening does
not occur; a condition true in most cases, which will be discussed below. For 
length and time scales larger than these estimates, the jet thins by an amount
determined by the balance of the two forces. Therefore, it is reasonable to
non-dimensionalize the experimental parameters with these scales to obtain
dimensionless heights and flow rates:%
\begin{equation}
H^{\ast}=\frac{H}{l_{\rm VG}}=H\left(  \frac{g}{\nu_{0}^{2}}\right)  ^{\frac{1}%
{3}}\label{Hstar}%
\end{equation}

\begin{equation}
Q^{\ast}=\frac{Q~t_{\rm VG}}{l_{\rm VG}^{3}}=Q\left(  \frac{g}{\nu_{0}^{5}}\right)
^{\frac{1}{3}}\label{Qstar}%
\end{equation}

At lower heights of fall the relevant scaling for the height is the
aspect ratio that is used in Eqs.~(\ref{coilbcklimit}) and
(\ref{foldbckllimit}):%

\begin{equation}
\epsilon=\frac{H}{2a_{0}}\label{epsilon}%
\end{equation}

This aspect ratio
will also be used as a dimensionless height.

The ratio of the elastic relaxation timescale $\lambda$ to $t_{\rm VG}$ gives rise to an
elasto-gravitational number. Non-Newtonian effects arise in the jet when the
stretch rate, driven by gravity and resisted by viscosity, becomes larger than
the relaxation rate of the wormlike micelles. Therefore, the
elasto-gravitational number characterizes the magnitude of these effects%
\begin{equation}
E_{g}=\frac{\lambda}{t_{\rm VG}}=\lambda\left(  \frac{g^{2}}{\nu_{0}}\right)
^{\frac{1}{3}}\label{Eg}%
\end{equation}

In a similar fashion, the radius of the jet at points along its stream
($i=0,1,2)$ (Fig.~\ref{schemes} and Fig.~\ref{all_scheme_cpycl})
can be scaled by the same length scale%
\begin{equation}
a_{i}^{\ast}=\frac{a_{i}}{l_{\rm VG}}=a_{i}\left(  \frac{g}{\nu_{0}^{2}}\right)
^{\frac{1}{3}}\label{radii}%
\end{equation}

In the folding regime, the dimensionless measured quantities that we report (i.e.~the frequency $f
$ and the amplitude $L$) are scaled in the same fashion:%

\begin{equation}
f^{\ast}=f~t_{\rm VG}=f\left(  \frac{\nu_{0}}{g^{2}}\right)  ^{\frac{1}{3}%
}\label{De}%
\end{equation}

\begin{equation}
L^{\ast}=\frac{L}{l_{\rm VG}}=L\left(  \frac{g}{\nu_{0}^{2}}\right)  ^{\frac{1}%
{3}}\label{LStar}%
\end{equation}

The Reynolds number defined in Eq.~(\ref{ReTimescales}) can also be rewritten
using these expressions as:%
\begin{equation}
Re=\frac{Q}{a_{0}\nu}=\frac{Q^{\ast}}{a_{0}^{\ast}}\label{Re}%
\end{equation}
\subsection{Reverse swell}

The reverse swell (Fig.~\ref{all_scheme_cpycl}(a)) is a very peculiar
feature of jets of wormlike-micellar fluids, in which the jet widens two to three times its initial radius near the plate, as compared to the thinnest part of the tail. For Newtonian fluids, the jet 
constantly thins, down to the coil region, beyond which the radius becomes constant. In
the case of CPyCl micellar jets, this reverse swell effect arises from the
non-Newtonian viscoelastic character of the fluid. As the fluid accelerates
under the effect of gravity, the fluid elements stretch, the wormlike micelles
become aligned and store elastic energy. When the local stretching rate becomes weaker,
because of the deceleration imposed by the presence of the plate, the
stretched molecules recoil, which leads to the observed lateral dilation. 
A noticeable reverse swell is therefore possible for a Weissenberg
number $Wi=\lambda(U_{1}-U_{0})/H=E_{g}(H^{\ast}-Q^{\ast}/a_{0}^{\ast2})$
greater than approximately unity. Note that this condition requires $H^{\ast}\geq Q^{\ast
}/a_{0}^{\ast2}$, which means that the fluid particles must be accelerated
during the fall. For practical purposes (in order to get a measurable reverse
swell), a more realistic condition is $H^{\ast}\gg Q^{\ast}/a_{0}%
^{\ast2}$, which simplifies the condition for reverse swell to
\begin{equation}
Wi=E_{g}H^{\ast}\gg1\label{Wi}%
\end{equation}

For example, the jet shown in Fig.~\ref{all_scheme_cpycl}(a) is
characterized by a large value of $Wi=11.3$ and exhibits significant reverse swell.
The reverse swell is strongly reminiscent of the die-swell phenomenon, and a derivation
similar to that given by Tanner \cite{Tanner} can be used to predict the amount
of swelling. The assumption is that no external force acts on the fluid
element over the length scale of recoil $h$, thus we neglect gravity on the
scale of $h\ll l_{\rm VG}$, as well as the reaction of the bottom plate. If $N_{1}$ 
is the first normal stress difference in the fluid just before the swell, $P$
the hydrostatic pressure in the jet, $I$ the identity matrix, and
$\alpha=a_{2}/a_{1}$ is the swell ratio, the force balance can be written as
\[
-PI+\left(
\begin{array}
[c]{ccc}%
3G & 0 & 0\\
0 & 3G & 0\\
0 & 0 & 3G+N_{1}%
\end{array}
\right)  \left(
\begin{array}
[c]{ccc}%
\alpha^{2} & 0 & 0\\
0 & \alpha^{2} & 0\\
0 & 0 & \frac{1}{\alpha}^{4}%
\end{array}
\right)  =0
\]

This corresponds to three equations with four unknowns. Eliminating $P$ among
these equations leads to
\begin{equation}
\alpha=\left(  1+\frac{N_{1}}{3G}\right)  ^{\frac{1}{6}}\label{swellRatio}%
\end{equation}

To proceed further, viscoelastic material elements will be considered as purely elastic during
the stretch process in the tail of the jet. This does not imply that the
elastic component of the total tensile stress in the jet is more important than the viscous
part, which would not be true, but only that the elastic part does not relax
during the time of fall, which is legitimate for moderate residence time in
the jet and $Wi\gg1$. We can therefore estimate $N_{1}$ from the tensile stresses
expected from rubber elasticity theory, given by%

\begin{equation}
N_{1}=3G\left(  \left(  \frac{a_{0}}{a_{1}}\right)  ^{4}-\left(  \frac{a_{1}%
}{a_{0}}\right)  ^{2}\right) \label{normalStress}%
\end{equation}

We have already assumed that $a_{1}\ll a_{0}$ (significant thinning),
therefore a first order approximation of Eq.~(\ref{normalStress}) combined with
Eq.~(\ref{swellRatio}) gives a swell ratio of
\begin{equation}
\alpha=\left(  \frac{a_{0}}{a_{1}}\right)  ^{\frac{2}{3}}\label{swellRatioDim}%
\end{equation}

In dimensionless form, and using the value for $a_{1}$ given by
Eq.~(\ref{a1Dimensional}), one finds that the swell ratio $\alpha$ should vary as
\begin{equation}
\alpha=\frac{a_{2}}{a_{1}}\sim\left(  a_{0}^{\ast2}\frac{H^{\ast}}{Q^{\ast}%
}\right)  ^{\frac{1}{3}}\label{swellRatioDimLess}%
\end{equation}

Equation (\ref{a1Dimensional}), which has the form $a_{1}\sim(Q/H)^{1/2}(\nu
/g^{2})^{1/6}$ when omitting the multiplicative constant, can be rewritten in
dimensionless terms as%
\begin{equation}
a_{1}^{\ast}\sim\sqrt{\frac{Q^{\ast}}{H^{\ast}}}\label{a1DimLess}%
\end{equation}

Combining Eqs.~(\ref{swellRatioDimLess}) and (\ref{a1DimLess}) leads to the
following scaling for the final radius:%
\begin{equation}
a_{2}^{\ast}\sim\left(  a_{0}^{\ast~2}\sqrt{\frac{Q^{\ast}}{H^{\ast}}}\right)
^{\frac{1}{3}}\label{a2DimLess}%
\end{equation}

Or in dimensional terms, this becomes%
\begin{equation}
a_{2}\sim\left(  a_{0}^{2}\sqrt{\frac{Q}{H}}\right)  ^{\frac{1}{3}}\left(
\frac{\nu_{0}}{g^{2}}\right)  ^{1/18}\label{a2Dim}%
\end{equation}

In addition, the reverse swell process takes place over a vertical length $h$, determined
by the balance between the characteristic speed of the downward flow in the
swollen region, which can be averaged to $Q/\pi a_{1}a_{2}$, and the
upward propagation of viscous effects, $\nu/h$. Using Eq.~(\ref{a2DimLess}), this
leads to the following dimensionless expression for $h$
\begin{equation}
h^{\ast}=\frac{h}{l_{\rm VG}}\sim\left(  \frac{a_{0}^{\ast4}}{H^{\ast2}Q^{\ast}%
}\right)  ^{1/3}\label{hEquation}%
\end{equation}

This can be written in dimensional terms as
\begin{equation}
h\sim\left(  \frac{a_{0}^{4}}{H^{2}Q}\right)  ^{1/3}\left(  \frac{\nu_{0}^{7}%
}{g^{2}}\right)  ^{1/9}\label{hDim}%
\end{equation}

Everything else being equal, the height $h$ over which the swell occurs scales 
as $\nu_{0}^{7/9}$. This means that viscoelastic fluids with low viscosities will display a sharp reverse swell, whereas for very viscous viscoelastic fluids it may be gradual and hard to discern. Another
way to say the same thing is to derive the slope in the swell region, $s=(a_{2}-a_{1})/h$,
using Eqs.~(\ref{a2Dim}) and (\ref{hDim}). In the limit of $a_{2}\gg a_{1}$ (a rather strong approximation) 
we obtain
\begin{equation}
s=(a_{2}-a_{1})/h\sim\frac{(HQ)^{1/2}g^{1/9}}{a_{0}^{2/3}\nu_{0}^{13/18}%
}\label{slope}%
\end{equation}

Equation (\ref{slope}) is of limited validity because the assumption
$a_{2}\gg a_{1}$ is not true in most cases, however, it underlines that
experimentally noticeable reverse swell requires a fluid with a viscosity as
low as possible. Since elasticity is important (from the condition
(\ref{Wi})), strongly elastic fluids such as wormlike micellar solutions 
are ideally suited for observing reverse swell, as opposed to very viscous viscoelastic fluids such as
Boger fluids or weakly elastic fluids such as commercial shampoos. 
Larger values of $h$ also mean that an increased time in the swell is
available for the macromolecules to relax, which may invalidate the purely
elastic recovery assumption used to derive Eq.~(\ref{swellRatioDim}). As a result,
it is not possible at the present time to be certain whether very viscous 
elastic fluids (such as Boger fluids) are not prone to reverse swell at all, or if
it is simply not experimentally noticeable. For example, Chai and
Yeow \cite{1988Chai} studied the shape of jets of a Boger fluid and found a small
widening at the base, although the bottom boundary condition was somewhat 
different from here: the jet was falling straight into a pool of the same
liquid and was not subject to buckling. Testing the mechanism described in this section
would require using another wormlike micellar fluid (different from CPyCl) but with
comparable features of significant elasticity, especially large elastic
component of extensional stress, and moderate or low viscosity.

\subsection{Scaling laws for folding dynamics}

Next, we seek to investigate the folding motion of CPyCl wormlike micellar jets. The goal is
to understand why this in-plane jet motion (never observed for cylindrical jets
of Newtonian fluids) is possible for this shear-thinning viscoelastic fluid. In
addition, we derive scaling laws for the folding amplitude $L$ and frequency
$f$ that will be compared to experimental measurements.

\subsubsection{Mechanism of folding}

The mechanism of folding of CPyCl wormlike micellar solution has different
roots from Newtonian coiling. Even when the jet is pushed sideways,
experimental observations show that it remains relatively straight (Fig.~\ref{detailMechanism}(a)), whereas Newtonian jets get bent and twisted over
a significant height from the bottom plate. This is allowed by the
shear-thinning properties of the fluid, which limit the shear stress in the
curved region, as shown in Fig.~\ref{detailMechanism}(c), and it is likely
that some variant of shear banding takes place in this region. The shear stress $\eta
(\dot{\gamma})\dot{\gamma}$ in the curved region is bounded by the plateau stress $\eta
_{0}/\lambda$, regardless of the curvature, whereas it scales with the
curvature squared in the Newtonian case (Eq.~(\ref{viscousTorque})). The
point of contact of the jet with the lower layer of fluid can therefore be in
line with the centerline of the jet, and move at the same pace as the rest of
the jet. Since it is not twisted the jet does not coil and follows a straight motion.

As the jet translates further and further sideways, the weight of the inclined
jet tends to provide a restoring force directed towards the vertical axis. 
This creates a bending torque within the jet, and it is resisted by a viscous torque that appears
when the lower part of the jet bends (Fig.~\ref{detailMechanism}(d)). When
the lateral displacement reaches its maximum amplitude $L$, the gravitational torque becomes
larger than the viscous torque, and the jet buckles back toward the center
line. Such a situation is shown in Fig.~\ref{detailMechanism}(b). The jet
then makes a new contact point with the layer of liquid, continues its
movement because of the steady incoming flow of new fluid from above, 
and the process repeats itself. 

In order to derive scaling laws for the amplitude $L$ and frequency $f$ of
folding we have to analyze what happens at the extremal point of this
motion. This analysis is similar to the scaling laws developed in
Skorobogatiy and Mahadevan (2000) \cite{2000MahaVS} for viscous sheets (i.e.
planar jets), except that here, the fact that the jet is straighter, changes
 the relevant scales.
\subsubsection{Connection between amplitude and frequency}

 In order to connect the frequency of folding to the
amplitude we need to obtain an equation for conservation of volume similar to
Eq.~(\ref{volumeConservation}). In contrast to the coiling jet, shown in Fig.~\ref{allGraphsMechanism}(a), 
the radius of the jet during the folding motion is not
constant in the bent part, as can be seen in Fig.~\ref{detailMechanism}(a). 
The small horizontal
velocity $U_{3}$ can be roughly estimated as the horizontal component of
$U_{2}=Q/\pi a_{2}^{2}$: it is approximately zero when the jet is vertical,
and largest when it approaches $L$. Taking an average over one period, we find that this velocity
scales as $U_{2}a_{2}/L$. As a result $f$ and $L$ are linked by
\begin{equation}
fL\sim U_{2}\frac{a_{2}}{L}\Rightarrow f\sim\frac{Q}{L^{2}a_{2}}%
\label{volumeConservationCPyClDim}%
\end{equation}

Various regimes exist depending on which force dominates in the jet. The
possible driving forces are either viscosity (V), gravity (G), or inertia (I). The
resisting force is always viscous. Table \ref{AllScalingLaws} summarizes the
scaling laws for frequency as a function of $H^{\ast}$ and $Q^{\ast}$ for the coiling motion in 
viscous Newtonian fluids, folding motion in viscoelastic fluids, and the scaling laws for the 
CPyCl solutions obtained via experiments. In the subsequent analyses the dominant force in each regime will be identified via the appropriate subscripts V, G, or I to the frequencies and amplitudes.
\subsubsection{Viscous regime}

Analogous to Newtonian fluids, at low heights of fall, the driving force for folding
is the viscous stress in the fluid near the nozzle. As in the case of viscous coiling,
the range of heights of fall for viscous folding will be limited by both the 
buckling height of the column and transition to gravitational folding. 
In the cases where viscous folding occurs, no gravitational
stretching occurs, so $a_{0}=a_{1}=a_{2}$. In this regime the jet amplitude is
geometrically constrained by the nozzle-plate distance $H$, which means that
\begin{equation}
L_{\rm V}\sim H\label{ViscAmplCPyCl}%
\end{equation}

The expression for the viscous folding frequency is found using Eq.~
(\ref{volumeConservationCPyClDim}):
\begin{equation}
f_{\rm V}\sim\frac{Q}{H^{2}a_{0}}\label{ViscFreqCPyCl}%
\end{equation}

In the viscous regime the motion of the jet is therefore totally constrained
by the external parameters of the experimental setup, rather than by the fluid properties.

\subsubsection{Gravitational regime}

For larger heights of fall, the driving torque is the gravitational torque
acting on an arm given by the extremal lateral displacement $ L$ This torque scales as%
\begin{equation}
T_{\rm driving}\sim\rho gHa_{1}^{2}L_{G}\label{GravTorqueCPyCl}%
\end{equation}

At the maximum amplitude $L$ the jet falls backward, bending on the length
scale $l_{\rm VG}$, typical of the opposing influences of gravity and viscosity. The
typical curvature of the jet just at that moment is of the order of
$\kappa=1/L$. This leads to local shearing deformation within the curved region of the column, 
with viscous stresses developing between the outer region (of larger velocity) and
the inner region. In this situation, the viscous stress $\tau_{xz}$ is the
direct analog of the elastic stress in beam bending: it vanishes in the middle
of the jet and increases linearly outwards throughout the cross-section. Following this
analogy in the fashion developed in \cite{2000MahaVS}, the stress is found to be:
\begin{equation}
\tau_{xz}=\eta_{0}\dot{\gamma}\sim\eta_{0}x\dot{\kappa}\sim\eta_{0}\frac
{x}{L_{\rm G}}\frac{Q}{a_{1}^{2}l_{\rm VG}}%
\end{equation}

Note that at the very onset of bending, the liquid is not sheared, therefore
the zero-shear viscosity is used. As expected, the viscous force $d^{2}%
F_{\rm V}=\tau_{xz}dxdy$ integrated over the jet cross-section vanishes. The
elementary viscous torque, $\delta^{2}T_{\rm V}=x\tau_{xz}dxdy$, nevertheless,
remains non-zero after integration. Approximating the cross-section to a
square, the resulting torque is
\begin{equation}
T_{\rm resisting}\sim%
{\displaystyle\int\limits_{-a_{1}}^{a_{1}}}
{\displaystyle\int\limits_{-a_{1}}^{a_{1}}}
x\tau_{xz}dxdy=\eta_{0}\frac{a_{1}^{2}Q}{l_{\rm VG}L_{\rm G}}\label{ViscTorqueCPyCl}%
\end{equation}

The balance between the two torques gives the scaling law for the amplitude of
folding%
\begin{equation}
L_{\rm G}\sim\left(  \frac{\nu_{0}Q}{l_{\rm VG} g H}\right)  ^{1/2}%
\label{GravAmplCPyClDim}%
\end{equation}

In dimensionless form, we obtain the following expression
\begin{equation}
L_{\rm G}^{\ast}\sim\left(  \frac{Q^{\ast}}{H^{\ast}}\right)  ^{1/2}%
\label{GravAmplCPyClDimLess}%
\end{equation}

The expression for $a_{2}$ in Eq.~(\ref{volumeConservationCPyClDim}) is found
using Eq.~(\ref{a2DimLess}), which leads to a dimensionless expression for the
frequency in this regime:%
\begin{equation}
f_{\rm G}^{\ast}\sim\frac{\left(  Q^{\ast5}H^{\ast}\right)  ^{1/6}}{a_{0}%
^{\ast2/3}L_{G}^{\ast2}}\label{VolumeConservationCpycl2}%
\end{equation}

Eliminating the amplitude using Eq.~(38), we can derive the following expression for the dimensionless folding frequency in the gravitational
regime:%
\begin{equation}
f_{\rm G}^{\ast}\sim\frac{1}{a_{0}^{\ast2/3}}\left(  \frac{H^{\ast7}}{Q^{\ast}%
}\right)  ^{1/6}%
\end{equation}

The transition between viscous and gravitational regimes takes place for
$\beta=Ha_{0}\sqrt{g/6Q\nu}>\pi/2$. For the CPyCl 100 wormlike micellar
solution (described in detail below), with $a_{0}=1.25$ mm and $Q=3$~mL/min, this
is equivalent to $H=$ $3.3$ cm. Once again, at low flow rates, the jet may
transition directly from steady flow to gravitational folding.

\subsubsection{Inertial regime}

In contrast to the inertial regime for coiling (\cite{1998MahaVJ}, \cite{2004RibeVJ}), the type of inertia that drives folding is not the
centrifugal (rotational) inertia but the linear, axial inertia. The fluid in
the jet tends to travel along a vertical path, which provides a restoring
force returning the folded part of the jet toward the center. The inertial
force per unit length is $\rho Q^{2}/a_{1}^{2}$, which leads to a torque that
scales as
\begin{equation}
T_{\rm driving}\sim\frac{\rho L_{\rm I}Q^{2}}{a_{1}^{2}}%
\end{equation}

In this regime the length of the bent region is much smaller than for the
gravitational regime, and scales with $L$ rather than $l_{\rm VG}$. This leads to
a viscous torque that scales as
\begin{equation}
T_{\rm resisting}\sim\eta_{0}\frac{a_{1}^{2}Q}{L_{\rm I}^{2}}%
\end{equation}

This leads to a folding amplitude that scales as%
\begin{equation}
L_{\rm I}\sim\left(  \nu_{0}\frac{a_{1}^{4}}{Q}\right)  ^{1/3}%
\label{InertiaAmplCPyClDim}%
\end{equation}

In dimensionless terms and using Eq.~(\ref{a1DimLess}), this can be written in the form%
\begin{equation}
L_{\rm I}^{\ast}\sim\left(  \frac{Q^{\ast}}{H^{\ast2}}\right)  ^{1/3}%
\label{InertiaAmplCPyClDimless}%
\end{equation}

The volume conservation condition, Eq.~(\ref{VolumeConservationCpycl2}), is once again used to obtain the
scaling law for the folding frequency:%
\begin{equation}
f_{\rm I}^{\ast}\sim\frac{H^{\ast3/2}Q^{\ast1/6}}{a_{0}^{\ast2/3}}%
\label{InerFreqFolding}%
\end{equation}

The inertial regime is expected to appear at large heights, of the order of
heights required to have $\rho Q^{2}/a_{1}^{2}\sim\rho gHa_{1}^{2}$; with
CPyCl 100 fluid parameter values, that height corresponds to $H\approx9\nu_{0}^{2}/(gl_{VG}^{2})=26.5$ cm. At 
these heights, thermal effects due to the evaporative cooling of water can 
also become very important and dramatically alter the viscosity. For this reason, no
systematic quantitative measurement has been conducted in this regime.%


\section{Experimental procedure}

\subsection{Fluid formulations}

The main focus of the study is the behavior of jets of wormlike
micellar solutions. In addition, we use a silicone oil as a reference fluid to facilitate 
qualitative comparison with a viscous Newtonian fluid.


The type of fluids we use for this study are solutions of CPyCl and
NaSal in brine (100 mM of NaCl salt solution). Following the study in
\cite{CABERCPyCl}, a fluid in the desired range of viscosity and elasticity is
obtained with the concentrations [CPyCl] = 100 mM, and [CPyCl]:[NaSal] = 2:1.
For the sake of comparison, two other fluids were prepared with the same brine
and the same [CPyCl]/[NaSal] ratio, but [CPyCl] = 75 mM and [CPyCl] = 150 mM,
respectively. The three fluids will be denoted as CPyCl 100, CPyCl75, and
CPyCl150, respectively, in the rest of the study. CPyCl and NaSal were obtained in dry form from
MP Biomedicals, and Sigma-Aldrich, respectively.


As a comparison fluid with Newtonian properties, we used a silicone oil T41
from Gelest Inc.
\subsection{Rheological characterization}

\subsubsection{Shear Rheology}

The linear viscoelastic tests were performed on all fluids on a stress-controlled ARG2
rheometer, in a cone-plate geometry of 40 mm diameter, at 22.5$^{o}$C. We
obtain the values of important fluid parameters such as the zero-shear rate
viscosity $\eta_{0}$ and the characteristic relaxation time $\lambda$, which
are summarized in Table \ref{RheologyData2}. The CPyCl solutions exhibit a
strong viscoelastic behavior, which is well characterized by small-amplitude
oscillatory experiments at low frequencies.  A single-mode Maxwell model fits the data well as
shown in Fig.~\ref{allGraphsRheology}(a). The relaxation time $\lambda$ in this
model is directly related to the characteristic time of the stress relaxation
processe in wormlike micellar solution, as shown in Eq.~(\ref{LambdaMicelles}). 
All the CPyCl
solutions are also strongly shear-thinning (Fig.~\ref{allGraphsRheology}(b)), which translates into a critical stress or plateau shear stress 
(in Fig.~\ref{allGraphsRheology}(c), we show the example of CPyCl 100). This plateau
stress corresponds to a stress level that is sufficient to break the weak 
intermolecular bonds holding the surfactant molecules in the wormlike
micelles, leading to their breakup in smaller aggregates. 
This apparent strong
shear-thinning, with $\eta\sim 1/ \dot{\gamma}$, can also indicate the 
onset of shear banding, as predicted by Spenley and co-workers \cite{1993Spenley}, 
and verified by Berret and co-workers \cite{1994Berret}. As we show in 
Fig.~\ref{allGraphsRheology}(b), below a critical shear rate of the order of 
$\dot{\gamma}_{c}\simeq1/\lambda$, CPyCl solutions show a plateau or 
zero-shear-rate viscosity (reported in Table \ref{RheologyData2}).

Another non-Newtonian effect is the first normal stress difference that arises
because the shear flow tends to deform and align the wormlike micellar network, 
which leads to a streamline tension resulting in a normal stress difference \cite{DPL}. 
In Fig.~\ref{allGraphsRheology}(c), we show that the first normal stress
difference increases approximately linearly with the shear rate over the measured range. 
At low shear rates ($\dot{\gamma}_{c}\ll1/\lambda$), where the first normal stress 
difference is expected to scale quadratically with the shear rate \cite{DPL}, the measured value are below the sensitivity threshold of the ARG2 rheometer.

\subsubsection{Extensional rheology}

 In Fig.~\ref{allGraphsCaber}(a), we show a sequence of snapshots of a 
typical CaBER experiment conducted on a CPyCl 100 solution. The diameter of
the plates is 6 mm, and they are initially separated by 1.2 mm. A step-strain is
imposed, pulling the plates apart in 50 ms to 4.8 mm, which represents a Hencky
strain $\epsilon=\ln(4.8/1.2)=1.4$.

As we show in Fig.~\ref{allGraphsCaber}(b), the three solutions behave in a
qualitatively different manner in this experiment. CPyCl 75 has such a small
viscosity that it almost immediately enters a regime of elasto-capillary
exponential thinning. At the other end of the spectrum, CPyCl 150 is almost
gel-like, and the liquid filament is sustained for an extended period of time.
Breakup eventually occurs, but usually not at the mid-plane \cite{CABERCPyCl},
thus the typical analysis used in CaBER is not applicable. The CPyCl 100 fluid
shows the most interesting behavior with strong non-Newtonian effects.

The first part of the thinning is dominated by a balance between capillary and
viscous effects. The mid-plane diameter decreases linearly in time. As this
diameter decreases, the extension rate increases, until it starts triggering
non-Newtonian effects. The wormlike micelles become increasingly aligned by
the extensional flow, building up elastic stresses, which lead to
extension-thickening. 
In Fig.~\ref{allGraphsCaber}(c), we show the first normal stress difference $\tau_{zz}-\tau_{rr}$ as a function of time (scaled by $\lambda$) for the three fluids. The main feature exhibited by each fluid is the strong extension thickening as seen from the rapid rise in the first normal stress difference beyond the Newtonian regime.
More extensive and detailed studies 
of capillary thinning experiments with wormlike micellar solutions are 
discussed by \cite{CABERCPyCl} and \cite{CaberMiller}.
\subsection{Temperature dependence}

The rheological properties of wormlike micelles solutions are highly temperature dependent, 
for two reasons. The first reason is that  the two characteristic timescales of the relaxation processes,
$\lambda_{r}$ and $\lambda_{b}$ evolve with
temperature. Here, $\lambda_{r}$ refers to the reptating motion
of the chain segments and $\lambda_{b}$ represents a thermally-activated
breakup process. Everything else held constant, both decrease exponentially
with temperature. The second source of temperature dependence is the variation of the
characteristic length of the micelles, which are dynamic structures. Their
average length depends exponentially on the thermally-activated processes of
association/dissociation of the surfactant molecules at the ends of the
micelles \cite{Cates}. This affects $\lambda_{r} $ through a power law
dependence according to the reptation theory of de Gennes \cite{DeGennes}: if
we denote the characteristic entanglement length of the micelles as $L_{e}$, $\lambda_{r}$
scales as $L_{e}^{3.4}$. The temperature dependence of each timescale 
contributes towards the temperature dependence of $\lambda$, the Maxwell model fit parameter for the small angle oscillatory stress data, through the connection given by Eq.
(\ref{LambdaMicelles}).

The temperature dependence of $\lambda$\ can be fitted with an Arrhenius
equation of the form
\begin{equation}
\lambda(\rm T)=\lambda_{\rm T}=\lambda_{\rm Tref}\exp\left(  \frac{\Delta H}{R}\left(
\frac{1}{T}-\frac{1}{T_{\rm ref}}\right)  \right) \label{Tdependency}%
\end{equation}

The ratio $\lambda_{\rm T}/\lambda_{\rm Tref}$ is called the shift factor, denoted by $a_{\rm T}$. In Fig.~\ref{AllGraphsTemp}(a), we
show that the temperature dependence of the shift factors $a_{\rm T}=\lambda
_{\rm T}/\lambda_{\rm Tref}$ with $T_{\rm ref}=21.5%
{{}^\circ}%
$C follows the Arrhenius model with a very large activation energy. This in
turns leads to a large temperature dependence of the viscosity of the fluid.

As a first approximation, the viscosity of a viscoelastic fluid such as CPyCl
100 is related to its elastic modulus and longest relaxation time by $\eta
_{0}=G_{0}\lambda$. Provided that the length of the micelles is long enough
for them to entangle, the elastic modulus depends mostly on the surfactant
concentration; $G_0 \sim \nu_{e} k_{b} T$, where $\nu_{e}$ is the number density of the entanglements. Rubber elasticity theory suggests that $G_{0}$ varies only
linearly with temperature \cite{Larson}. An estimate of $G_{0}$, the norm of
the complex modulus $|G^{\ast}|~=\sqrt{G^{\prime~2}+G^{\prime\prime~2}}$, is
indeed found to be approximately constant with temperature, as can be seen on
the right-hand side of Fig.~\ref{AllGraphsTemp}(b). As a result, the
zero-shear-rate viscosity of the solution varies exponentially with
temperature. This is found to be true experimentally, with an exponential
factor of the same order as the elastic relaxation time. The viscosity of the CPyCl 100 solution drops from 25.2 Pa.s at 20$%
{{}^\circ}%
$C to 9.3 Pa.s at 25$%
{{}^\circ}%
$C. This means that in order to get meaningful experiments with this type of
fluid, the temperature must be either carefully controlled, or measured for
each experiment in order to correct and re-evaluate the fluid properties in accordance with
Eq.~(\ref{Tdependency}).
Another affect of the temperature can be seen in capillary breakup and
filament stretching experiments, as well as during the breakup of the jet
falling on a plate at very low flow rates. In these cases, the liquid,
initially transparent at ambient temperature (22-23$%
{{}^\circ}%
$C), becomes increasingly turbid. Direct temperature measurement with a thermocouple
immersed in the small mound of white liquid falling on the bottom plate shows
that its temperature can be as low as 16$%
{{}^\circ}%
$C. This suggests that the turbidity is caused by the Krafft
transition, i.e. the precipitation of surfactant from the solution when
the temperature becomes lower than a critical temperature called the Krafft
temperature \cite{IUPACKrafft}. 
The Krafft
temperature can be measured either by visual observation of turbidity or by
the drastic change in most physical properties when the precipitation occur.
Figure~\ref{AllGraphsTemp}(b) offers one example of such a measurement, with
the abrupt change of the slope of the norm of the complex modulus $|G^{\ast
}|~$ as a function of $T$ in a small angle oscillatory shear test. This leads to an estimate of the 
Krafft temperature for CPyCl 100 of $18.0^{o}$C, which is consistent with
estimates from direct visual observations.

The cooling in itself is likely due to the evaporative cooling of water, which
has a strong effect on the temperature of thin filaments because of their
divergent surface area to volume ratio. The smaller radius also increases the local
capillary pressure, which in turn makes the chemical potential of the solvent
higher and accelerates evaporation. Another possibility connecting evaporation
and visual appearance of turbidity could simply be that the filament dries,
leaving only solid surfactant. These extreme cooling and drying
effects tend to appear only the limit of very large jet heights and low flow
rates (for example $H=25$ cm, $Q=2$ mL/min). The evaporation-driven cooling and concentration
increase leads to a large increase in viscosity when the residence time of a
fluid particle in the jet becomes of the order of the typical time for the
temperature change to take place. In other words, as the height of fall
increases and the flow rate decreases, the variability in the local viscosity increases.

The silicone oil shows the typical features of a purely Newtonian fluid, with
a constant viscosity. It is a fairly viscous fluid, with values of viscosity similar to those of CPyCl 100, which make it suitable for jetting experiments.%

\subsection{Experimental setup for jet analysis}

Two setups were used for the jetting studies in this research. The first setup
simply involves the fluid being pumped by a syringe pump to a nozzle, from where the jet falls
onto a plate. Direct observation of the jet profile and dynamics is sufficient
to obtain the flow regime, and a video camera is used for quantitative
measurements. The second setup is a different view configuration, where we use a laser projected along the fluid column, which helps in precise evaluation of the motion of the jet.
Advantages and disadvantages of each method are discussed below. We also discuss the precise definition and possible variants of the bottom plate condition.

\subsubsection{Direct observation}

The fluid is pumped by a Stoelting syringe pump (Stoelting 53130) with a controlled flow rate. The fluid is placed in a 60 mL syringe, and a flexible plastic tube is attached to it. The other end of the tube has a nozzle attached to it, through which the fluid exits as a jet. The nozzle is attached to a vertical-axis linear stage, and the liquid falls onto a
plate below the nozzle (see Fig.~\ref{allGraphSetup}(a)). The plastic tube is
approximately 20 cm long, which implies that the residence time of fluid particles in
the tube is between 10 and 20 seconds, greatly exceeding the relaxation times
listed in Table \ref{RheologyData2}. This ensures the relaxation of any stress
occurring at the exit of the syringe. We mostly report results obtained with a
circular nozzle of diameter $a_{0}=1.25$ mm, but the influence of the nozzle
radius has also been investigated with two additional sizes, $a_{0}=0.775$ mm
and $a_{0}=2.40$ mm. The motion of the jet was recorded using a
BlueFox digital videocamera at frame rates from 30 to 50 fps. Frequency
measurements were done by frame counting, and amplitude measurement were done
using the image analysis software ImageJ, using an image of a ruler as
reference. A high-speed videocamera (Phantom V from Vision Research) was 
also used to capture rapid phenomena such as jet breakup and the Kaye effect,
at frame rates from 500 to 800 fps.

\subsubsection{Trajectory tracking with laser}

A second setup was used to follow the trajectory of the jet, inspired from an
experimental technique used by Versluis \cite{2006Versluis} to study the Kaye
effect. A red He-Ne laser beam shines through the jet using a T-shaped nozzle and is
guided along the jet like in an optic fiber (Fig.~\ref{allGraphSetup}(b)).
The camera records the position of the beam spot through the transparent 
bottom plate, using a mirror at a 45$%
{{}^\circ}%
$ angle. An image analysis code then evaluates the precise position of the
spot. The technique allows for the quantitative understanding of the trajectory
of the jet in folding motion, including tracking over many periods that
reveals the stability of the folding regime. The shortcomings of this technique are that the
liquid jet is an imperfect waveguide, and therefore allows the laser beam to be transmitted 
through the free surface of the jet when the angle of incidence is too large,
which happens at large amplitude. Conversely, at very small amplitudes of the jet motion changes in the position of the laser spot are smaller than the size of the spot and hence difficult to quantify. 

\subsubsection{Discussion of the bottom-plate condition}

Two questions regarding the bottom-plate conditions must be addressed to
ensure that the problem is correctly defined. First of all, the jet does not
fall directly on the plate, but on a thin layer of fluid that is covering the
plate. We are only interested here in steady state regimes, for which the jet
falls on a layer of fluid, as opposed to the initial transients when the jet falls on the clean plate. The thickness of this layer is ``chosen" by the fluid, to
balance the viscous stresses between the upper, free layer and the no-slip
plate-fluid interface, with the incoming flow rate. This thickness is taken into 
account in the measurement of the height of fall, which is defined as the
distance between the fluid layer and the nozzle. The spreading fluid is either
allowed to collect into in a secondary reservoir when it reaches the end of the plate,
or the plate is cleaned before being the next run.

The second point to note is that other geometries beyond the flat plate could be envisioned, which may be relevant for industrial applications. The jet could fall on a bath of the same or of a
different fluid, on an inclined plane, or on a curved surface. Different plate
sizes could also influence how fast the fluid layer
drains into the secondary reservoir. In the present study, we are interested
in planar geometries. The drainage mechanism does not play a significant
role, because of the fact that the layer thickness is taken into account for
the measurement of the height of fall, and the short duration of the
measurement. The liquid bath, in the end, is the only alternative that could
lead to dramatically different behavior such as air
entrainment \cite{2004Lorenceau}, at least for large incoming speed.
Nevertheless, for the moderate jet speeds involved in most parts of the
parameter space studied in this paper, all phenomena of interest occur 
on timescales shorter than the spreading time for the viscous
fluids used. As a result, the jet motion studied in our experiments always
occurs on a small mound of fluid that has not spread completely.

\subsection{Experimental results}

We first begin by a qualitative description of the different jetting regimes
involved, which are mapped onto the parameter space described earlier. 
This helps in understanding what experimental conditions are required for the novel regimes
of CPyCl solution jets to take place. Then, we investigate quantitatively the
reverse swell and folding phenomenon, and the results are compared to
theoretical predictions.

\subsubsection{Regime maps}

We map the different jetting regimes for the Newtonian silicone oil (see Fig.~\ref{regime_diagram_T41_CPyCl}(a)) and for
the wormlike micellar solution CPyCl 100 (see Fig.~\ref{regime_diagram_T41_CPyCl}(b)). The maps presented in Fig.~(\ref{regime_diagram_T41_CPyCl}) are drawn in the dimensionless ($Q^{\ast}=Q(g/\nu^{5})^{1/3}%
$, $\epsilon=H/2a_{0}$) parameter space. The scaling of $Q^{\ast}$ allows for comparisons with different fluid viscosities, which is especially
important for comparing experiments with CPyCl solutions conducted at different
ambient temperatures. The choice of $\epsilon$ as a relevant
dimensionless height is due to the fact that the predictions for jet buckling 
(Eqs.~(\ref{coilbcklimit}) and (\ref{foldbckllimit})) use this geometrical parameter.
Nevertheless, the other dimensionless height $H^{\ast}=H(g/\nu^{2})^{1/3}$
becomes more relevant when the focus is on large-height behavior.


\begin{description}
\item \bf Steady jet
\end{description}

In this regime, the jet is steady over time, and the fluid spreads
at a uniform rate on the plate. This regime occurs when the jet is not buckled, which means at low heights of fall, and is
similar for all the test fluids. This flow state is commonly known as the stagnation slow. 

\begin{description}
\item \bf Non-continuous jets
\end{description}
In this regime, which occurs at low flow rates and moderate heights, the jet is non-continuous. This
means, for the Newtonian silicone oil, that the fluid will \textit{drip} from the
nozzle rather than form a jet, or form a jet that periodically breaks under
the effect of surface tension. For viscoelastic fluids with large extensional
viscosity such as the CPyCl solutions, it may also mean the formation of
persistent thinning filaments between the nozzle and the plate with beads of
fluids periodically sliding down the filament. This transient situation is a
direct application of CaBER experiments, and is also reminiscent of the
beads-on-a-string \cite{BeadsOnString} and gobbling
phenomena \cite{ClasenGobbling}.

\begin{description}
\item \bf Coiling
\end{description}
Coiling is the typical mode of motion for buckled jets of Newtonian fluids. 
As reported in the regime map of
Fig.~\ref{regime_diagram_T41_CPyCl}(b), jets of CPyCl solutions can also
show coiling, although the range of experimental parameters for which it
happens is more limited than in the Newtonian case. In addition, the shape and
curvature of coiling jets of CPyCl jets (see Fig.~\ref{all_scheme_cpycl}(c)) is
different from their Newtonian counterpart (see Fig.~\ref{schemes}(c))
and the equations for Newtonian coiling may not apply.

\begin{description}
\item \bf Folding
\end{description}

The folding regime of CPyCl jets is studied in detail because it is
qualitatively different from the dynamical motion of Newtonian jets. In 
Fig.~\ref{allGraphsFolding}, we show two snapshots of the same jet of CPyCl
100 at different instants during a folding period. In Fig.~\ref{allGraphsFolding}(a), we 
show the jet when it is almost vertical, whereas in 
Fig.~\ref{allGraphsFolding}(b) the lateral jet displacement is at its maximum amplitude $L$ and is
about to fall back. In addition, in Fig.~\ref{allGraphsFolding}(c), we show an
example of the jet trajectory obtained by the laser tracking system. One can
see the oscillatory motion is confined primarily to a fixed plane, thus justifying the
concept of folding. We can also see the events of coiling, that occurs when
the fluid builds up a secondary mound from which the jet tends to be
deflected. Nevertheless, after two coils, the jet returns spontaneously to its folding
motion, because this regime is more stable under these specific experimental conditions.

\begin{description}
\item \bf Bistable regime
\end{description}

In this regime the jet either coils or folds, depending on the history of the
flow and the boundary conditions at the bottom plate. Small perturbations such as the
presence of a heap of fluid on the bottom plate can trigger the switch between the regimes. In
this case both regimes are stable and the system can be forced from one to another. In the same way coiling jets of Newtonian fluids can also display bistable states over a limited range of impact height or flow rate. 

\begin{description}
\item \bf High-flow-rate ductile failure
\end{description}

At large flow rates and high extension rate, micellar fluids tend to break in
a rubber-like ductile failure \cite{2003Rothstein}. This happens when the
weight of the fluid column pulling on a particular cross-section of the jet,
usually close to the nozzle, becomes larger than the stress the micelles can
sustain. As a result, the micelles break locally, leading to a local weakening
of the jet, in turn leading to rupture of the entangled micelles. This
creates a fracture pattern reminiscent of rubber failure in solid mechanics,
as can be seen in Fig.~\ref{rupture}. The dynamics of this ductile pinch-off  for a 
micellar network have been considered by Cromer et al. \cite{Cromer}.

\subsection{Comparison of the regime maps}

Figure \ref{regime_diagram_T41_CPyCl}(a) shows the regime map for the 
viscous  Newtonian oil (silicone oil T41),
for which only three regimes are typically observed: steady jetting, dripping, and
coiling. Note that sub-classification of coiling in viscous, gravitational and
inertial regimes is not taken in account here, since it is hard to determine
without actually measuring the coiling frequency. This can be done by frame
counting using the setup of Fig.~\ref{allGraphSetup}(a) and has already been performed in the studies by Ribe \cite{2004RibeVJ}. The buckling transition at
moderate flow rates is consistent with Eq.~(\ref{coilbcklimit}) (dashed line). At
large flow rates, the increased compressive stress favors buckling, and the
transition happens at a lower height.

The dripping-jetting transition (solid line) is observed to obey the
scaling law $\epsilon\sim Q^{\ast1.7}$. Using
Eq.~(\ref{Qstar}) and Eq.~(\ref{epsilon}), we find that the flow rate required to maintain a
continuous jet increases with height of fall and scales as
\begin{equation}
Q^{{}}\sim H^{1/1.7}\simeq H^{0.6}\label{NewtDripScalingExp}%
\end{equation}

Figure~\ref{regime_diagram_T41_CPyCl}(b) shows the corresponding experimental regime map
for CPyCl 100. Several features can be noted in comparison to the Newtonian
diagram. Some features  are somewhat similar to the Newtonian regime map; a 
stable axisymmetric jet is maintained at low heights of fall, and a non-continuous
jet develops at low flow rates. Most of the parameter space is occupied by
time-dependent buckled jets. Nevertheless, significant changes are also noticeable compared to Newtonian fluids, such as the existence of the folding regime, the ductile failure at large flow
rates, the coexistence of bistable folding and coiling regimes, and the different slope
of the dripping-jetting transition. Measurements with CPyCl 75 and 150 show a behavior very similar
to CPyCl 100 and the corresponding regime maps are not shown here for brevity.

\subsection{Experimental scaling laws for the regime transitions of CPyCl
solutions}

In this section we report detailed investigations of the transitions between different regimes. The experiments are 
for different CPyCl solutions and experimental conditions (flow rate, height of
fall, and nozzle size). In Fig.~\ref{allGraphTransitions}, we
present the data for one transition each, with respect to the relevant
dimensionless parameters at that transition. The horizontal axis is always $Q^{\ast}=Q(g/\nu
^{5})^{1/3}$, which scales for the effect of viscosity and allows a direct
comparison between the different solutions. The vertical axis is either
$\epsilon=H/2a_{0}$ or $H^{\ast}=H(g/\nu^{2})^{1/3}$, depending on
which variable allows for a better collapse of the data set. The column aspect ratio $\epsilon$ is 
expected to be more relevant at low heights, whereas $H^{\ast}$ should be the more relevant variable at larger
heights, when gravitational thinning takes place. The nozzle radius was
corrected to take into account the die-swell observed experimentally \cite{Tanner}, which is
especially important at large flow rates and small nozzle radii.

\begin{description}
\item \bf Buckling limit
\end{description}

In Fig.~\ref{allGraphTransitions}(a), we show that the jet buckles and starts folding when the aspect ratio
$\epsilon \approx 4.8$. This value is in good agreement with the 
prediction given by Eq.~(\ref{foldbckllimit}), $\epsilon=4.81$ (solid line). There is a strong hysteresis between the unbuckled and folding regimes when
the height of the fall is varied continuously. This  explains in part the scatter of the
data. Nevertheless, the transition happens for $3<\epsilon<8$ when
$Q^{\ast}$ \ is varied across three orders of magnitude. Newtonian fluids
buckle at a higher value of $\epsilon$, $\epsilon \simeq 7.66$ as
given by Eq.~(\ref{coilbcklimit}).
\begin{description}
\item \bf The first Folding-Coiling transition
\end{description}
Above the limit given by Eq.~(\ref{coilbcklimit}), the jet switches to the
azimuthal mode of instability and begins to coil. In Fig.~\ref{allGraphTransitions}(b), we show this transition from folding to coiling, which occurs at $\epsilon \simeq 7.66$. This is the limit at which Newtonian jets
usually buckle and start coiling. It immediately suggests that the analysis
that led to the limits (Eq.~(\ref{coilbcklimit})) and (Eq.~(\ref{foldbckllimit})) for the
Newtonian fluids is still valid for the CPyCl solutions, although in the case
of CPyCl solutions the folding mode is stabilized by mechanisms that do not
exist in the Newtonian case. Experiments show that the transition between the
folding and coiling occurs over a somewhat broad range of heights for which
the two modes alternate; the switching between the two being triggered by
random events such as the jet impacting a heap of liquid that has not
evenly spread out. Cruickshank \cite{1988Cruickshank} also noted this problem
in the narrow parameter range where folding was observed with Newtonian
fluids. It is interesting to note that when the ambient temperature is closer
to $25%
{{}^\circ}%
$C, so that the viscosity is lower, the heaps of liquid are less pronounced,
and the frequent switching is replaced by a bistable region (shown in Fig.~\ref{regime_diagram_T41_CPyCl}(b)), leading to a large hysteresis in the
transition. Overall, there is some scatter around the predicted value of
$\epsilon \simeq 7.66$ for this transition, with $4<\epsilon<12$,
but once again this remains valid over three orders of magnitude for $Q^{\ast
}$.

\begin{description}
\item \bf Second Folding-Coiling transition
\end{description}

When the height of fall is raised even more, a second transition occurs, from coiling back to folding, with an
even more pronounced bistability than the first transition (see Fig.~\ref{regime_diagram_T41_CPyCl}(b)). We
measured the transition from coiling to bistable folding and coiling. In other
words, starting from a well-developed coiling state and for a given flow rate, 
we raised the height until we saw the first events of folding. For this transition, the parameter
that allowed the best collapse of the data was the dimensionless height of
fall $H^{\ast}$ rather than the aspect ratio $\epsilon$. In contrast to the measurements
of the previous two boundaries, the critical height for this transition varies
with the flow rate. As the flow rate increases over three orders of magnitude, the height at which the transition occurs also increases, but as a power-law. Experimentally, the scaling is close to
$H^{\ast}\sim Q^{\ast1/3}$, as shown in Fig.~\ref{allGraphTransitions}(c).


\begin{description}
\item \bf Jet rupture at large flow rate: ductile failure
\end{description}

At very large flow rates, the entangled wormlike micellar network cannot sustain the axial
stresses anymore and break \emph{en masse}, leading to the solid-like ductile failure
of the jet. The flow rate required to
observe jet rupture decreases with height, and scales as $Q^{\ast}\sim
H^{\ast-1/1.54}=H^{\ast-0.65}$, as shown in Fig.~\ref{allGraphTransitions}(d). There is an overlap between the coiling and
jet rupture zones, for which the jet has enough time to coil a few times
before breaking. Since the weight of the column of fluid is the driving force
for this mode of jet breakup, it happens at large flow rates
and heights of fall, in the upper-right corner of the stability diagram shown in Fig.~\ref{regime_diagram_T41_CPyCl}(b).

\begin{description}
\item \bf Dripping-Jetting transition at low flow rates
\end{description}

We now describe the dripping to jetting transition boundary, which occurs at low flow rates and large heights. Dripping occurs when the height is large enough at a given flow rate such that a continuous jet can not be maintained. Instead of Newtonian dripping and capillary breakup, the non-continuous regime
for CPyCl 100 is observed to consist of long, thin filaments, which are prevented from
breakup by elastic forces. A significant difference with the Newtonian regime
map is that the transition between the non-continuous and continuous jet
happens at a constant flow rate, $Q^{\ast}>$ $Q_{\min
}^{\ast}$, for all heights, above a minimum height (see Fig.~\ref{regime_diagram_T41_CPyCl}(a) and (b), non-continuous regime). Since it is the elasticity of the 
entangled micellar network that holds the thread and
prevents breakup, a continuous jet is sustained above a threshold flow rate.

\subsubsection{Measurements of Jet Radius}

For the wormlike micellar fluids, we have shown that the jet radius increases near the plate and the minimum radius is achieved at a certain height above the plate. 
In order to measure the amount of swelling with respect to the experimental
parameters $H^{\ast}$ and $Q^{\ast}$, we use the setup described in Fig.
\ref{allGraphSetup}(a). We perform three sets of experiments, one with a
constant flow rate of $Q^{\ast}=7.1\times10^{-5}$, and two sets of complementary 
experiments with a constant height of fall, with $H^{\ast}=1.7$ and $H^{\ast}=3.4$. 
The relevant conjugate variable ($H^*$ or $Q^*$, respectively) was then slowly varied. The measurements are shown in Fig.~\ref{allGraphsRadius}. Here, $a_{1}^{\ast}=a_{1}(g/\nu^{2})^{1/3}$ is the scaled radius of the jet just above the swell, $\alpha=a_{2}/a_{1}$ is the swelling ratio, and $h^{\ast}=h(g/\nu^{2})^{1/3}$ is the scaled height of the swollen region. We compare the experimental values of these parameters to the 
theoretical predictions of Eqs.~(\ref{a1DimLess}),
(\ref{swellRatioDimLess}) and (\ref{hEquation}), respectively. In Fig.~\ref{allGraphsRadius}(a), we show the jet profile along with the definitions of different length scales. In Fig.~\ref{allGraphsRadius}(b), we show the dimensionless radius just above the swell. In 
Fig.~\ref{allGraphsRadius}(c), we show the swelling ratio and in Fig.~\ref{allGraphsRadius}(d), we show the height of the swollen region. The data for the different
experiments collapse well with the theoretical scaling. The only departure from theory is
the value of $h^{\ast}$ for one of the data sets, for which the onset of
swelling was difficult to detect.

\subsubsection{Jet dynamics}

The variations of folding frequency $f$ and amplitude $L$ with respect to
experimental parameters are presented in Fig.~\ref{allGraphsFrequency}.
Multiple series of experiments were performed to fully capture the folding
dynamics, in two sets. In the first set, three series of amplitude and
frequency measurements were made 
with a fixed flow rate and varying height. The
imposed flow rate was $Q=2$ mL/min for the first two series and $Q=5$ mL/min
for the third one. The series of tests were conducted at different ambient
temperatures, $22.5^{o}C$, $20.6^{o}C$, and $21.5^{o}C$, which
also affects the viscosity of the fluid, leading to three values of imposed
dimensionless flow rates: $Q^{\ast}=4.3\times10^{-5},$ $Q^{\ast}%
=7.1\times10^{-5}$, and $Q^{\ast}=1.7\times10^{-4}$, respectively. For each series, the flow
rate was held constant, while the fall height of the jet was varied from 1.2
to 20 cm ($0.4<H^{\ast}<6.7$). Note that amplitude data were difficult to
collect at low values of $H^{\ast}$ and are therefore reported over a smaller
range than the corresponding frequency data. In the second set of tests, four series of
experiments were performed with a fixed height of fall and varying flow rate.
The heights of fall were $H = 5, 6, 9$ and $10$ cm (with temperatures of
$22.5^{o}C$, $21.3^{o}C$, $21.8^{o}C$, and $22.5^{o}C$, respectively, and dimensionless
heights given by $H^{\ast} = 1.7, 1.9, 3.2, 3.4$, respectively). The flow
rate $Q$ was varied between 0.5 and 9 mL/min ($1.5\times10^{-5}\leqslant Q^{\ast
}\leqslant2.1\times10^{-4}$). All experiments were done using the CPyCl 100
fluid, because it is better suited for studying continuous jets as discussed earlier and
it helps in
eliminating other factors affecting the dynamics, especially fluid rheology.
The elasto-gravitational number for this fluid at $22.5^{o}C$ is $E_{g}%
=\lambda\left(  g^{2}/\nu_{0}\right)  ^{\frac{1}{3}}=12.4$. The high elasto-gravitational number implies that significant elastic effects are expected to occur at fall
heights when gravitational thinning is important.

Experimental results and the associated scaling laws are reported in Fig.~\ref{allGraphsFrequency}. In Figs.~\ref{allGraphsFrequency}(a) and (b), we show the frequency measurements,
while in Figs.~\ref{allGraphsFrequency}(c) and (d) we show the corresponding amplitude data. Panels (a) and (c) are the first
set of experiments with fixed flow rate, while (b) and (d) are the second set
with fixed height of fall. Data points for each set of experiments are collapsed using experimental scaling laws obtained from the other set iteratively until 
self-consistency is achieved. Eventually, the following experimental scaling laws for folding motion are found:
\begin{subequations}
\begin{align}
\text{For frequency, }f^{\ast} &  \sim H^{\ast0.66}Q^{\ast-0.19}%
\label{freqExpCpyCl}\\
\text{For amplitude, }L^{\ast} &  \sim H^{\ast0.50}Q^{\ast0.60}%
\label{amplExpCPyCl}%
\end{align}
\end{subequations}
where, as defined in the non-dimensionalization scheme, $H^{\ast}=H(g/\nu
^{2})^{1/3}$, $Q^{\ast}=Q(g/\nu^{5})^{1/3}$, $f^{\ast}=f(\nu/g^{2})^{1/3} $,
and $L^{\ast}=L(g/\nu^{2})^{1/3}$. 

We note that the power laws in Eqs.~(\ref{freqExpCpyCl}) and
(\ref{amplExpCPyCl}) are obtained from global regression of all experiments
taken together, whereas each series of experiments give scaling exponents that are different 
from the overall trend. The causes for these
variations are difficult to control and result from experimental conditions,
such as the effect of temperature on extensional viscosity, the effect of ambient
humidity on evaporation and cooling, or rheological aging of the solution.
These parameters are constant for a given series of experiments, but vary systematically from
series to series, therefore a statistical analysis based on the hypothesis of
stochastic deviation from a trend is not applicable. It is only possible to
give the range of scaling exponents obtained from different series, which are
provided in Table \ref{datarange}. Within this range of uncertainty,
theoretical prediction for the folding amplitude and frequency and
experimental measurement (all gathered in Table \ref{AllScalingLaws}) agree.%

\section{Conclusion}
The jetting dynamics of viscoelastic shear-thinning fluids
impacting on a plate are qualitatively different from their viscous analog. Cylindrical jets of CPyCl wormlike micellar
solutions falling on a plate from a sufficient height tend to exhibit a folding
transition with a periodic oscillating lateral motion. This contrasts with Newtonian
fluids for which this folding motion is only observed for planar sheets.
We can understand this
phenomenon by noting the pronounced shear-thinning of the
fluid, which allows a local drop in viscosity at the most curved part of the
jet, reducing the local viscous torque. While the Newtonian jets deal with this
viscous torque by twisting and coiling out of plane, wormlike micellar jets can remain 
planar
and fold backwards and forwards. Another novel feature of these jets of
wormlike micellar fluids is the widening or ``reverse swell" at the base. We have shown that this feature can be interpreted in terms of the extensional elastic stress, stored during
the stretching of fluid elements in the portion of the jet accelerated by gravity, and
partially recovered in the compressive part close to the bottom plate. We have
provided scaling laws for these two features that agree broadly with
Êexperimental measurements. In addition, we have documented the
locations of the different jetting regimes on dimensionless regime maps (see Fig.
10), and studied in more detail the transitions between the different regimes
for CPyCl micellar jets.

Many other classes of filling operations exploit jetting of complex fluids.
Surfactant-based fluids like Sodium Lauryl Ether Sulphate (SLES) are an
important ingredient in many consumer products including shampoos and liquid
detergents. Some other consumer products like conditioners, toothpastes, and
food products exhibit finite yield stresses. Future work will involve these
Êclasses of non-Newtonian fluids to examine the differences in their jetting
behavior. For example, in addition to coiling, fluids exhibiting yielding behavior tend
to show pronounced mounding \cite{EntovJNNFM}. In addition to using different
fluids, other possible extensions include a study of the jetting behavior in
confined geometries, replicating container filling process \cite{EntovJNNFM},
and the jetting behavior at higher speeds and consequently higher shear rates,
which are closer to actual industrial filling process in terms of the relevant
dimensionless operating parameters.

\section{Acknowledgements}
This research was supported by a gift from the Procter and Gamble Company.


\newpage
\section*{Figure Captions}

\noindent {\bf Figure~1:} Instabilities of a fluid jet impacting a plate. a) The jet remains
axisymmetric at low heights. b) Compressive forces in the jet lead to buckling
at a critical aspect ratio. c) Coiling jet. Panels a) and b) are jets of CPyCl, a wormlike micellar fluid, 
which is described in detail in the text. Panel c) is a silicone oil jet, a Newtonian fluid.

\noindent {\bf Figure~2:} Instabilities in the jet of a Newtonian fluid. a) Coiling jet of Newtonian silicone oil. b) Schematic view of the
coiling motion of an axisymmetric jet of Newtonian fluid.

\noindent {\bf Figure~3:} Schematic view of the assembly process of wormlike micelles.

\noindent {\bf Figure~4:} Jets of a wormlike micellar fluid. a) Close-up of the reverse swell for a jet of CPyCl 100 with $H=6$ cm
and $Q=0.5$ mL/min ($H^{\ast}=1.8 $, $Q^{\ast}=1.3\times10^{-5}$, $Wi=11.3$); the measured variables are also indicated. The black bar is approximately
1 cm. b) Snapshot and schematic view of the folding jet of CPyCl 100 with $H=6$ cm and $Q=3$ mL/min
($H^{\ast}=1.8$, $Q^{\ast}=8.0\times10^{-5}$). c) Snapshot of a coiling jet of CPyCl 100 with $H=3$ cm and $Q=5$
mL/min ($H^{\ast}=0.9$, $Q^{\ast}=1.3\times10^{-4}$).

\noindent {\bf Figure~5:} Different views of folding mechanism. a) A CPyCl jet at the
farthest position. b) Schematic view of the folding mechanism: the jet tends to
fall vertically under its own weight, which is resisted by a viscous torque. c)
At the farthest position of the oscillation, just at the onset of buckling of
the main part of the jet under the weight of the jet, the fluid is not
sheared, whereas the contact zone with the fluid layer is in highly
shear-thinning or shear-banding conditions. d) Close-up of the buckled region
of the jet, in which curvature induces shear stress $\tau_{xz}$.

\noindent {\bf Figure~6:} Rheological properties of micellar solutions. a) Small amplitude
oscillatory test of CPyCl 100, fit by a single-mode Maxwell model (similar fits
were found for CPyCl 75 and CPyCl 150). Here, $\lambda = 0.72$s, and $G_0 = 26.04$Pa. 
b) Steady shear viscosity of CPyCl 75
($\blacktriangle$), CPyCl 100 ($\blacksquare$) and CPyCl 150 ($\bullet$), as a
function of shear rate. c) Steady shear first normal stress difference as a
function of shear rate.

\noindent {\bf Figure~7:} Capillary Breakup Extensional Rheology (CaBER) experiments, with
plates of 6 mm diameter and an imposed Hencky strain of 1.4. a) Side view of
an experiment with CPyCl 100 ($\tau_{Break}=15.8$ s). b) Evolution of the
diameter with time scaled by the solution relaxation time measured in shear
and reported in Table \ref{RheologyData2}, for CPyCl 75, 100, and 150. Each
solution behaves in a qualitatively different fashion. c) First normal stress difference for the same three fluids.

\noindent {\bf Figure~8:} Temperature dependence of the viscometric properties of CPyCl 100 micellar solution. 
a) Shift factors with
$T_{ref}=294.65$ $K$. b) Krafft transition at $T=18^{o}C$ shown as change of
slope of the complex modulus $|G^{\ast}|~=\sqrt{G^{\prime~2}+G^{\prime
\prime~2}}$ as a function of $T$, in an oscillatory shear experiment at a
frequency of $\omega = 1 ~{\rm s}^{-1}$.

\noindent {\bf Figure~9:} Experimental setups. a) Experimental setup for regime diagram and
quantitative measurements. b) Experimental setup for trajectory
visualization.

\noindent {\bf Figure~10:} Experimental regime maps for silicone oil (Panel a) and CPyCl 100 (Panel b) in
the $\epsilon=H/2a_{0}$ and $Q^{\ast}=Q(g/\nu^{5})^{1/3}$ space. a) The
silicone oil shows only three behaviors in the ranges of heights and flow rates
investigated: steady jet ($\blacktriangle$), dripping ($\square$), and coiling
($\bullet$). b) CPyCl 100 also shows folding ($\blacksquare$), bistable coiling
and folding ($\varnothing$), elastic rupture ($\bigstar$). Solid lines are
guide to the eye for regime transitions, while the dashed lines are
Cruickshank's prediction for buckling transition (\ref{coilbcklimit}) and
(\ref{foldbckllimit}).

\noindent {\bf Figure~11:} Different views of the folding motion. a) Folding jet as it reaches the
central vertical position. b) The same jet at its
farthest lateral displacement, at the onset of buckling under the jet's weight. c) An
example of the trajectory of the laser spot (viewed from below), for a jet of CPyCl 100, from a height H = 11.4 cm ($H^{\ast}=3.86$), with a flow rate Q = 3 ml/min ($Q^{\ast}=1.06\times
10^{-4}$). The values of the Reynolds and Ohnesorge number Re = $10^{-3}$ and
$Oh^{2}=10^{-4}$ are typical of the jetting experiments in this paper.

\noindent {\bf Figure~12:} Successive views of the high-flow-rate rupture of a jet of CPyCl 100,
at $H^{\ast}=4.8$ and $Q^{\ast}=$ $2.7\times10^{-4}$ ($H=16$ cm, $Q=10$
mL/min). a) Onset of the ductile failure. All snapshots are separated by 
$\delta t = 24$ ms, which corresponds to $\delta t / \lambda = 0.033$. 

\noindent {\bf Figure~13:} The transitions between different flow regimes with CPyCl 100 and
$a_{0}^{\ast}$ = $~2.6\times10^{-2}$ ($\blacksquare$); CPyCl 75 and
$a_{0}^{\ast}$ = $~4.2\times10^{-2}$($\bullet$); CPyCl 100 and $a_{0}^{\ast}$
= $~4.2\times10^{-2}$($\blacktriangle$); CPyCl 150 and $a_{0}^{\ast}$ =
$~4.2\times10^{-2}$($\blacklozenge$); and CPyCl 100 and $a_{0}^{\ast}$ =
$~8.1\times10^{-2}$($\blacktriangledown$). a) The critical condition for
transition between steady axisymmetric flow and folding (buckling) at
$\epsilon=4.8\pm2$ b) Transition from folding to coiling when the
height of fall is increased from a low height for a given flow rate at
$\epsilon=7.6\pm1$ c) Transition from folding to coiling when the
height of fall is decreased from a large height for a given flow rate. d)
Appearance of jet rupture event when increasing flow rate for a given height
of fall. 

\noindent {\bf Figure~14:} Quantitative measurements of the dynamics in the tail. Here,  
$Q^{\ast}=7.1\times10^{-5}$, $0.4\leqslant H^{\ast}\leqslant6.8$
($\blacktriangle$); $H^{\ast}=1.7$, $2.8\times10^{-5}\leqslant Q^{\ast
}\leqslant3.2\times10^{-4}$($\bullet$); and $H^{\ast}=3.4$, $2.8\times
10^{-5}\leqslant Q^{\ast}\leqslant2.1\times10^{-4}$($\blacksquare$). a)
Definitions of the measured variables. For this jet $H^{\ast}=2.4$, $Wi=29.8$.
b) Dimensionless radius just above the swell. c) Swelling ratio $\alpha
=a_{2}/a_{1}$ d) Height of the swollen region.

\noindent {\bf Figure~15:} Quantitative measurements of folding properties, frequency (a and b)
and amplitude (c and d). The first set of experiments with fixed flow rate (a and
c): $Q=2$ ml/min ($Q^{\ast}=4.3\times10^{-5}$) ($\bullet$); $Q=2$ ml/min
($Q^{\ast}=7.1\times10^{-5}$) ($\blacksquare$); $Q=5$ ml/min ($Q^{\ast
}=1.7\times10^{-4})$ ($\blacktriangle$). The second set of experiments with fixed
height of fall $H=5$ cm ($H^{\ast}=1.7)$ ($\boldsymbol{\circ}$); $H=6$ cm
($H^{\ast}=1.9)$ ($\boldsymbol{\square}$); $H=9$ cm ($H^{\ast}=3.2)$
($\boldsymbol{\triangle} $); $H=10$ cm ($H^{\ast}=3.4)$
($\boldsymbol{\triangledown}$).

\clearpage

\newpage

\newpage
\begin{figure}[bthp!]
\begin{center}
\includegraphics[width=450pt]
{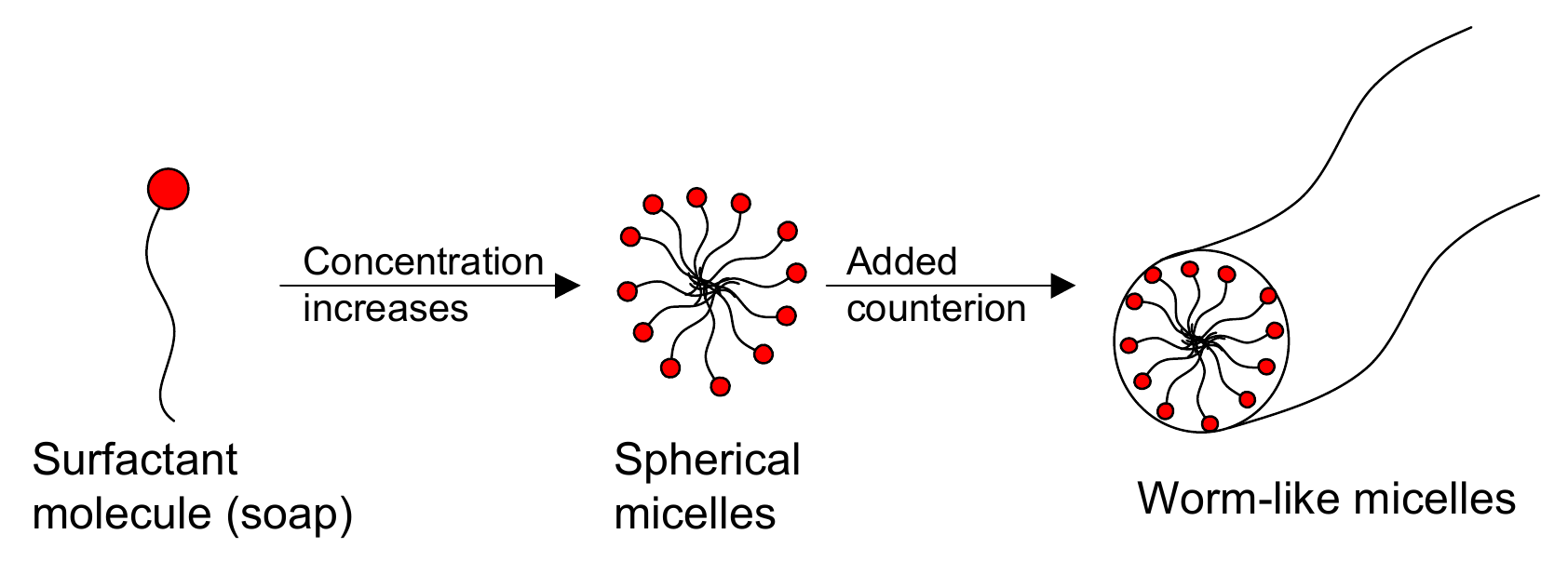}
\caption{}
\label{wormschemes}
\end{center}
\end{figure}

\newpage
\begin{figure}
[!ht]
\begin{center}
\includegraphics[width=400pt]
{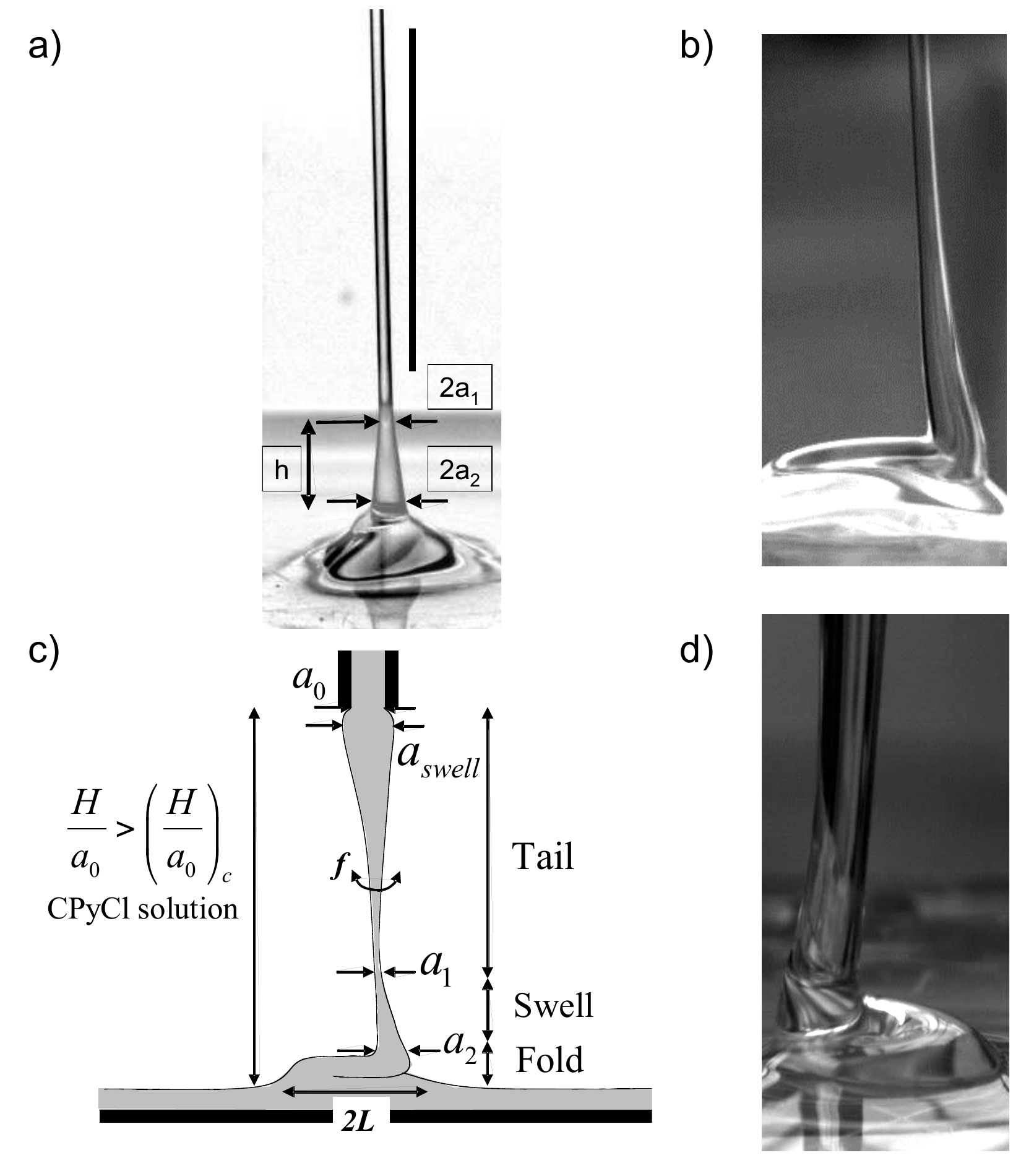}
\caption{}
\label{all_scheme_cpycl}%
\end{center}
\end{figure}

\newpage
\begin{figure}[!ht]
\begin{center}
\includegraphics[width=400pt
]
{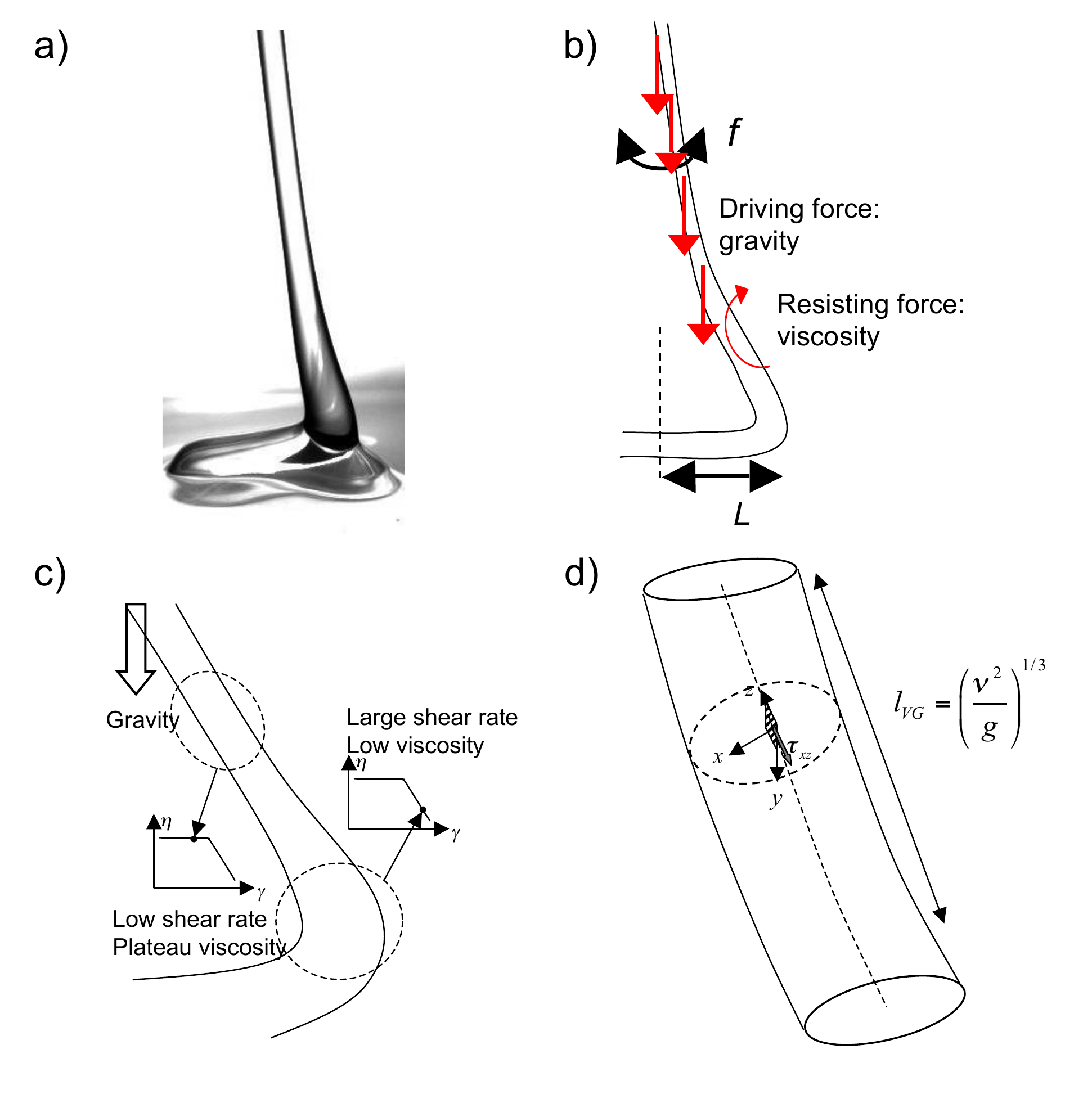}
\caption{}
\label{detailMechanism}
\end{center}
\end{figure}

\newpage
\begin{figure}
[bthp!]
\begin{center}
\includegraphics[width=425pt
]
{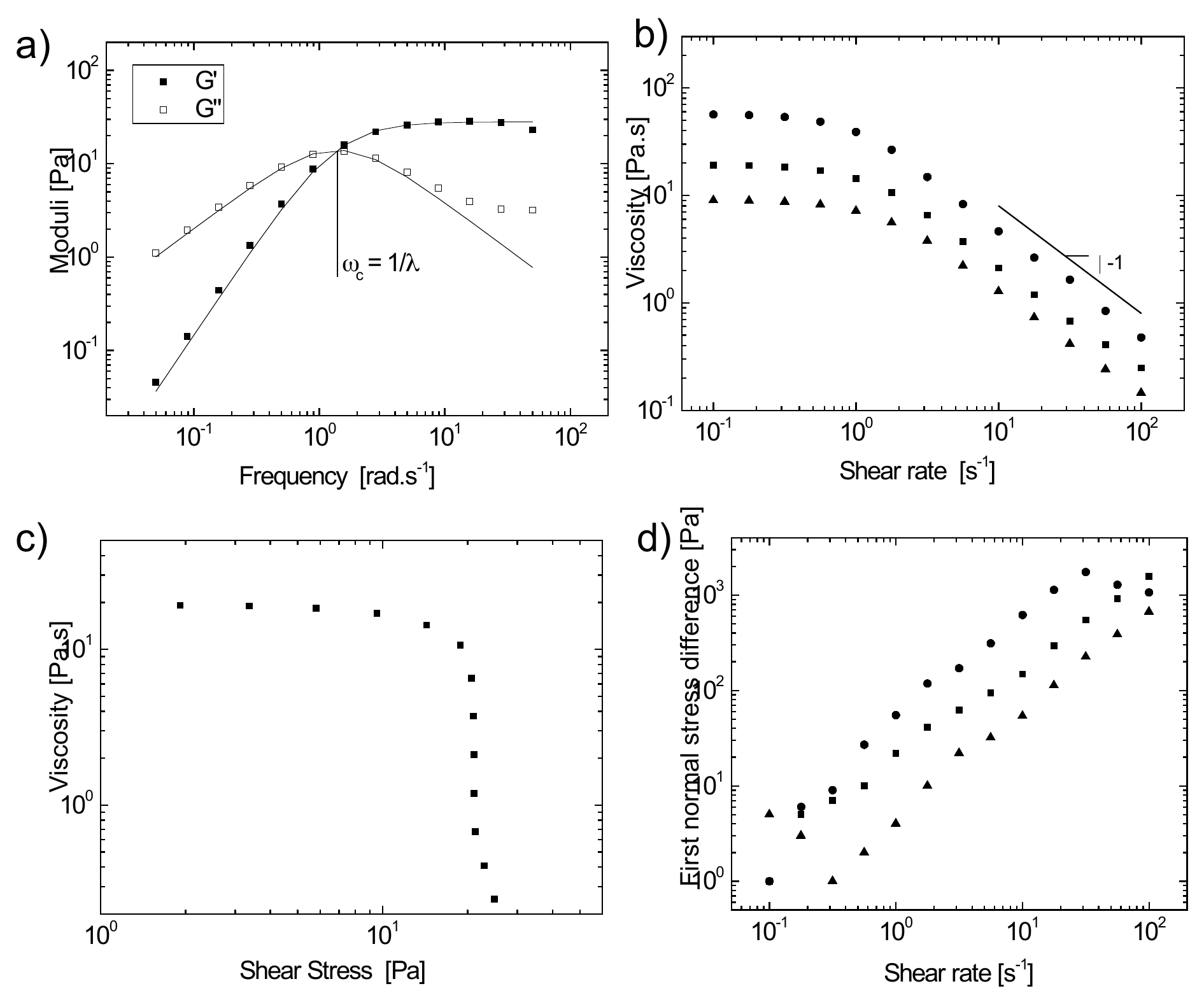}
\caption{}%
\label{allGraphsRheology}%
\end{center}
\end{figure}

\newpage
\begin{figure}
[bthp!]
\begin{center}
\vspace{1pt}\
\includegraphics[width=350pt
]
{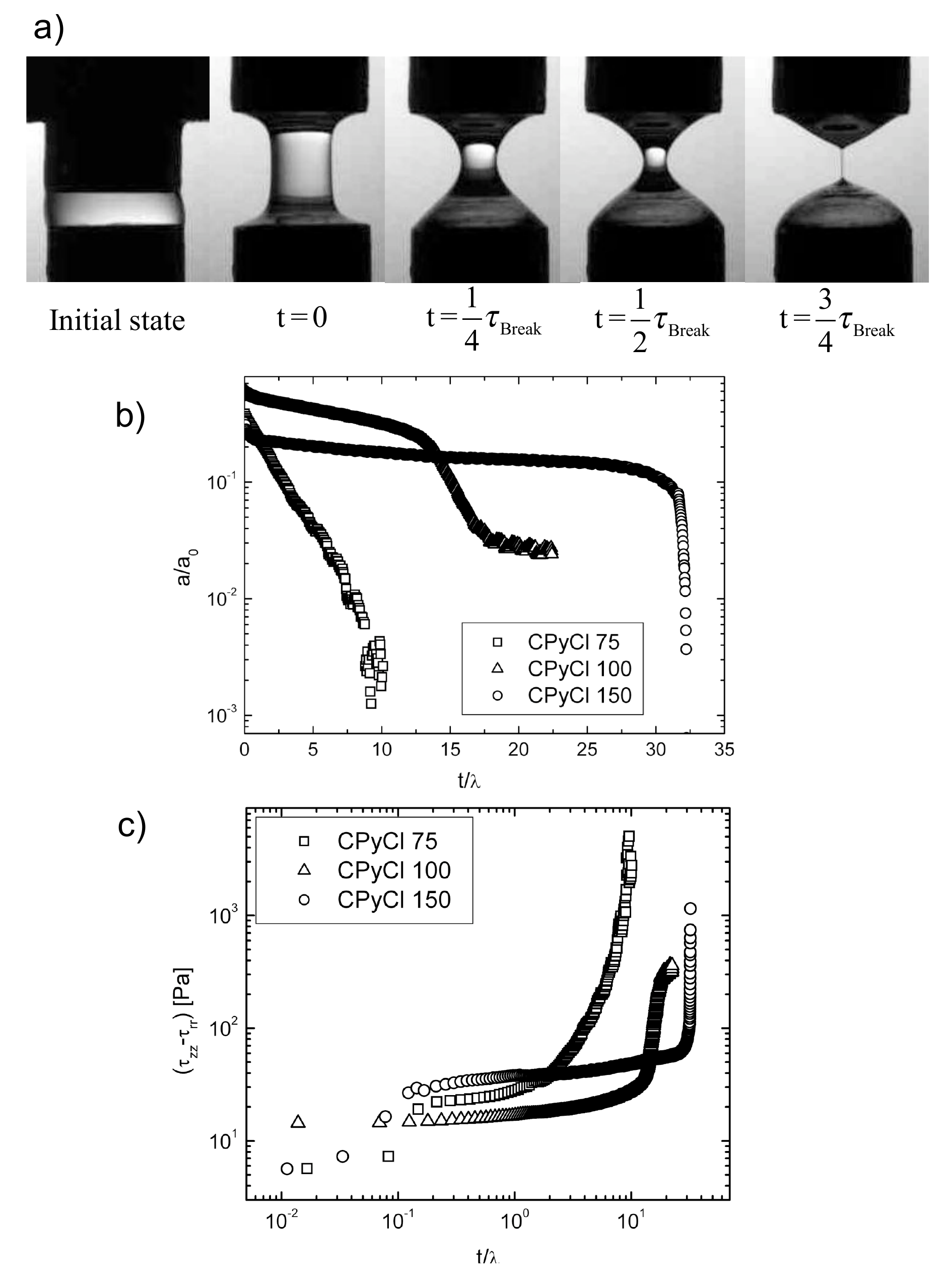}
\caption{}%
\label{allGraphsCaber}%
\end{center}
\end{figure}

\newpage
\begin{figure}
[bthp!]
\begin{center}
\includegraphics[width=425pt
]
{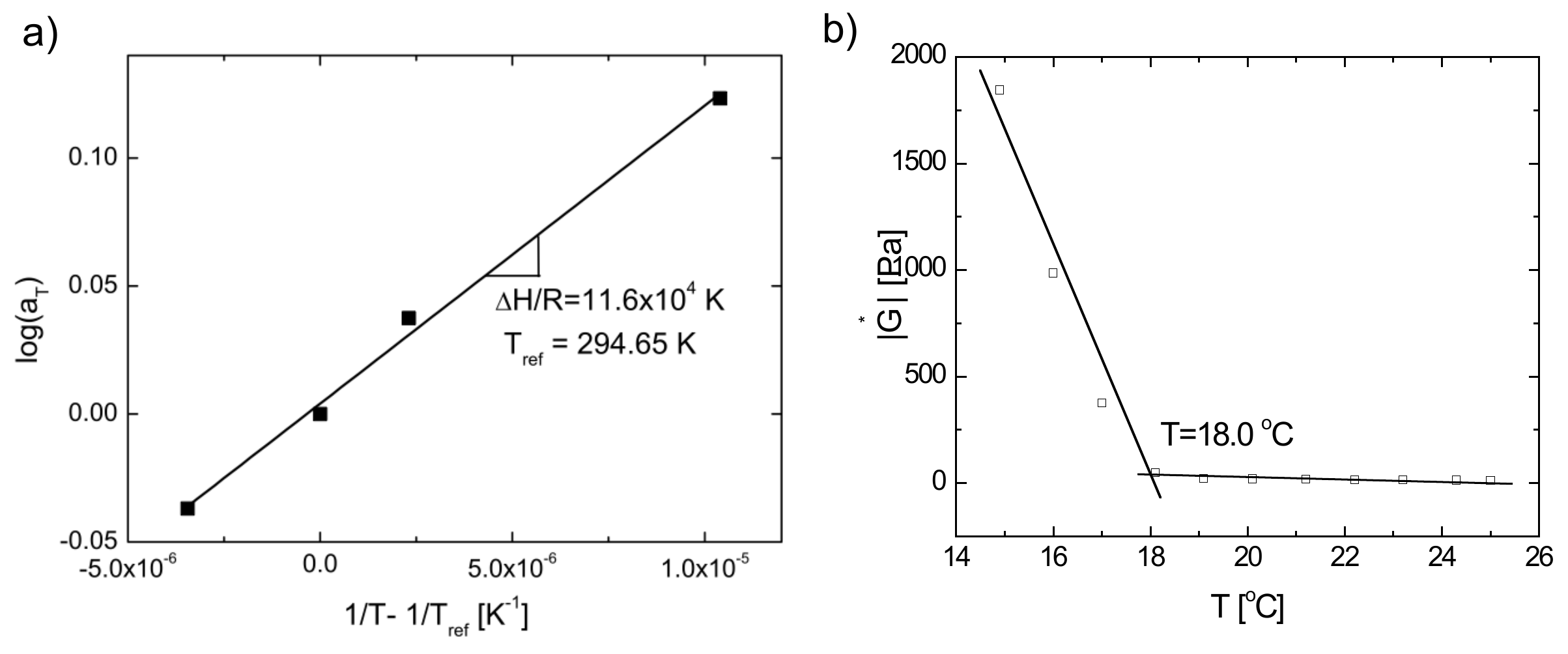}
\caption{}%
\label{AllGraphsTemp}%
\end{center}
\end{figure}

\newpage
\begin{figure}
[bthp!]
\begin{center}
\includegraphics[width=425pt
]
{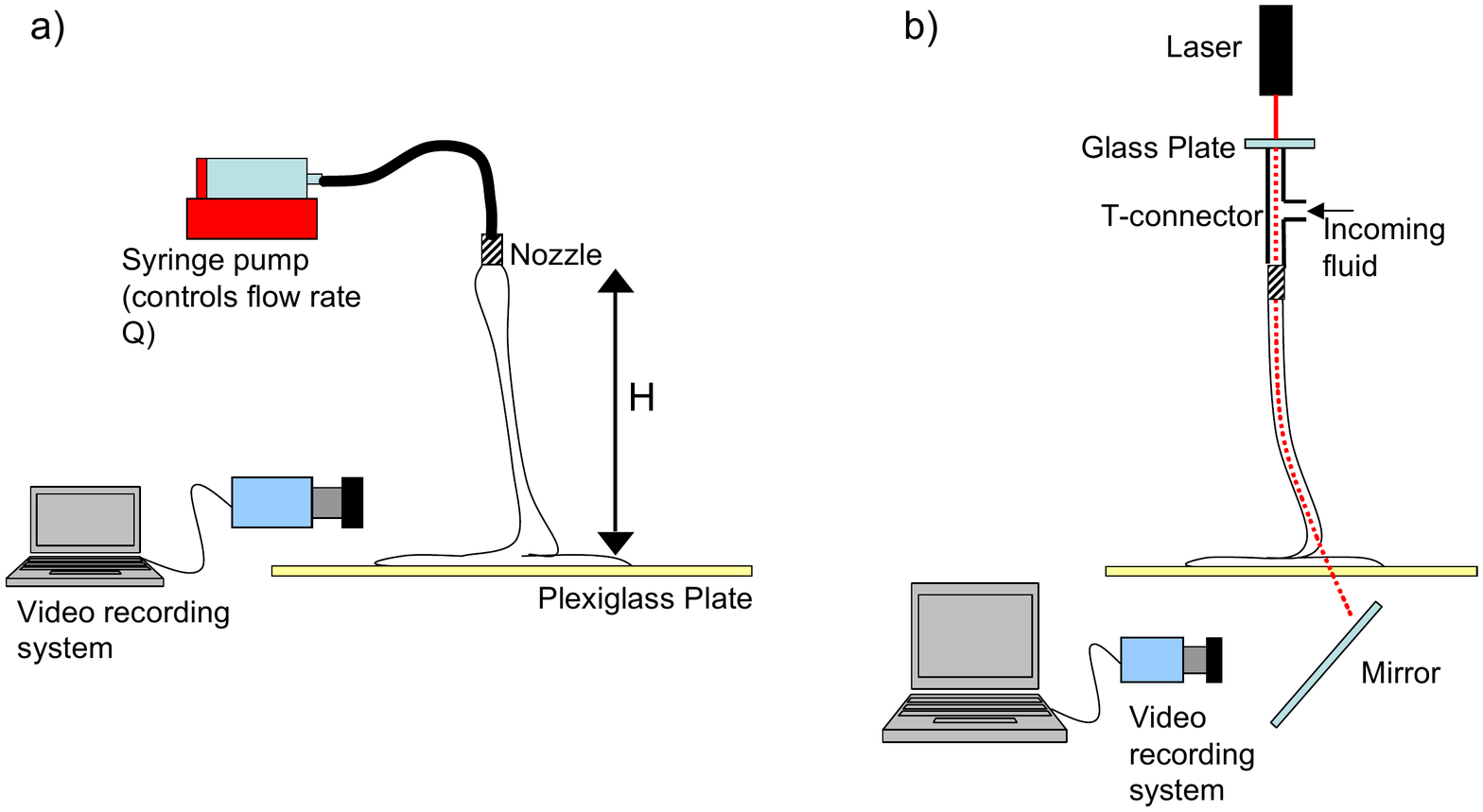}
\caption{}%
\label{allGraphSetup}%
\end{center}
\end{figure}

\clearpage
\begin{figure}
[bthp!]
\begin{center}
\vspace{1pt}
\includegraphics[width=350pt
]
{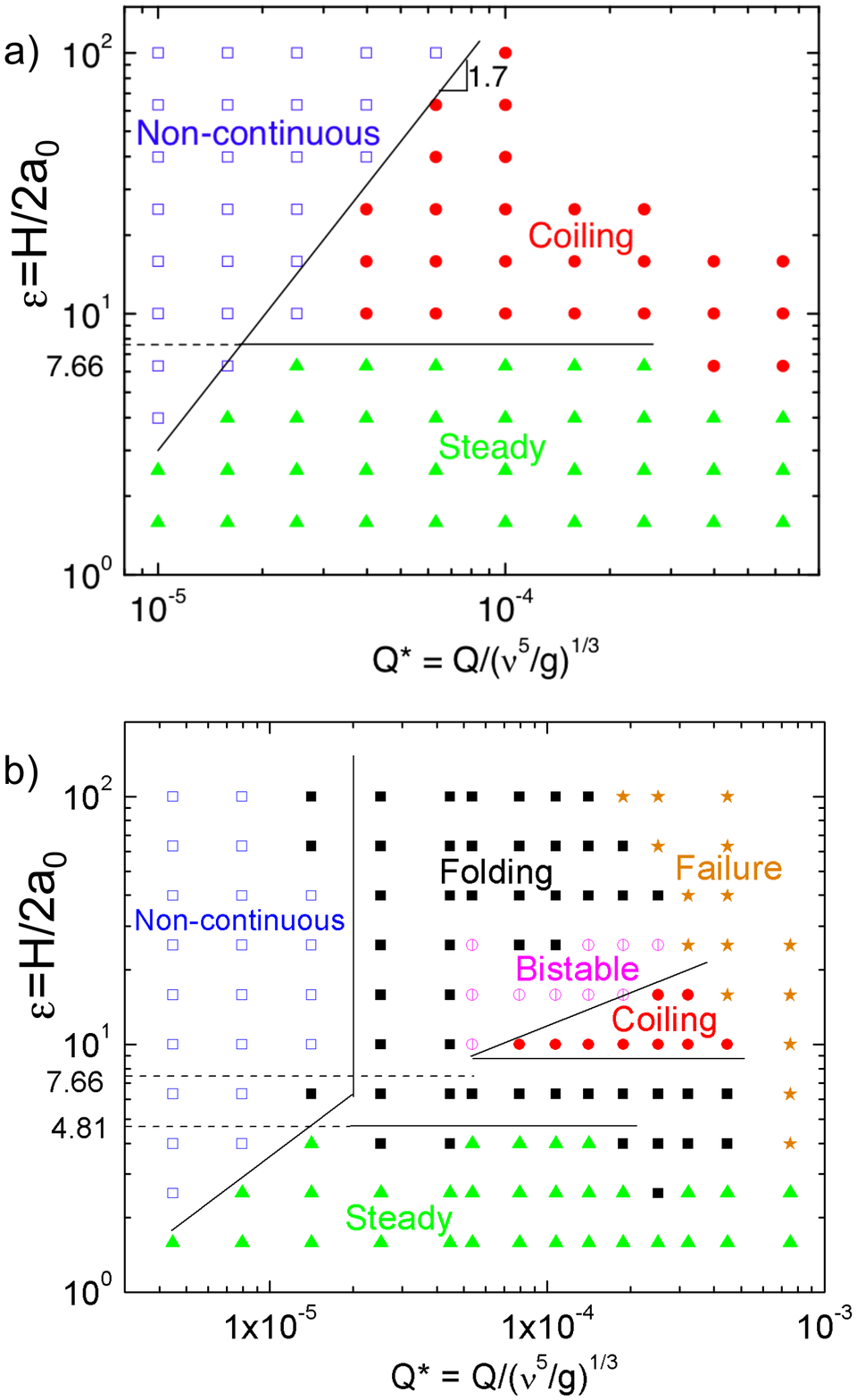}
\caption{}%
\label{regime_diagram_T41_CPyCl}%
\end{center}
\end{figure}

\clearpage
\begin{figure}
[bthp!]
\begin{center}
\includegraphics[width=425pt
]
{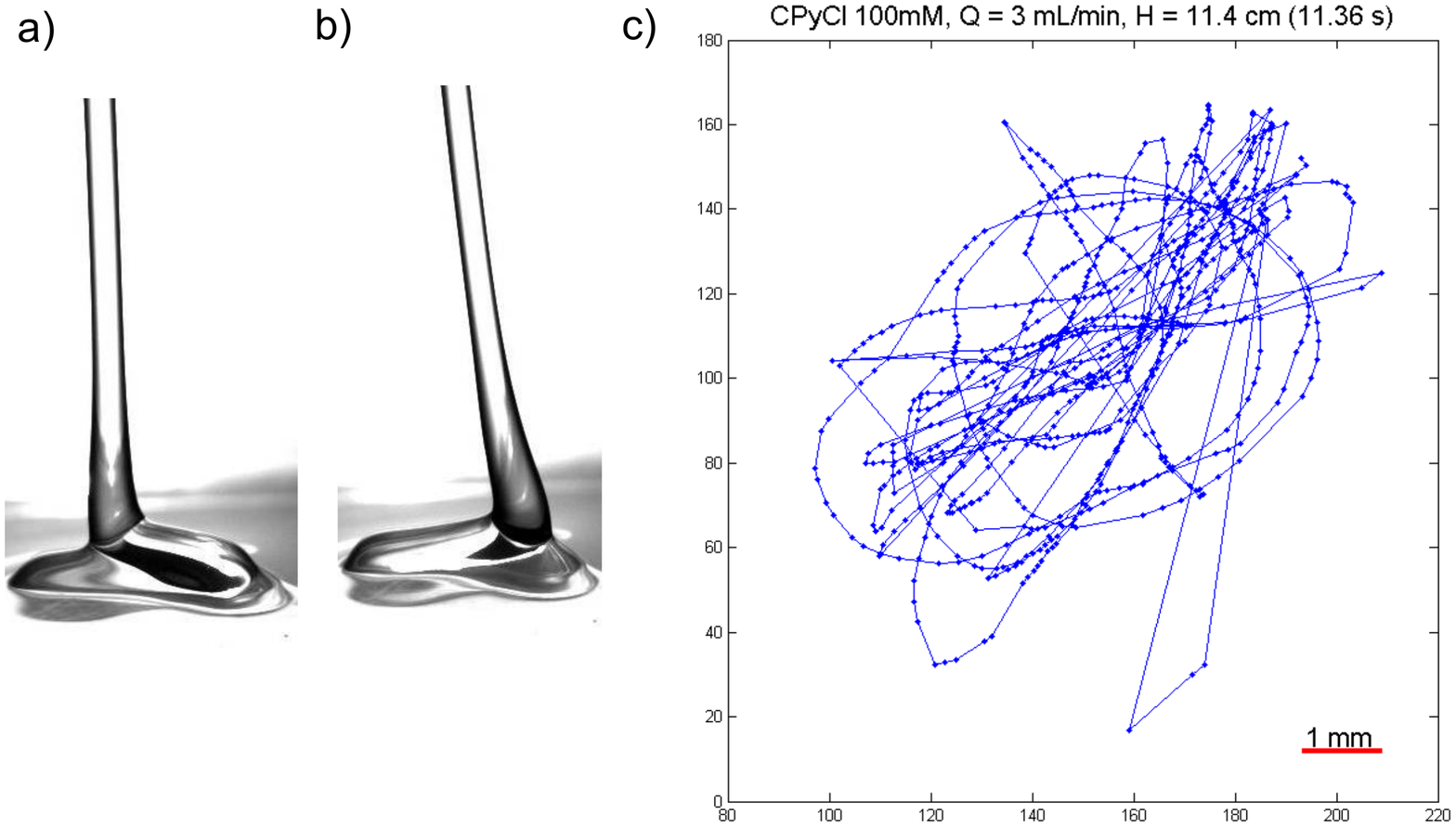}
\caption{}
\label{allGraphsFolding}
\end{center}
\end{figure}

\clearpage
\begin{figure}
[bthp!]
\begin{center}
\vspace{1pt}
\includegraphics[width=400pt
]
{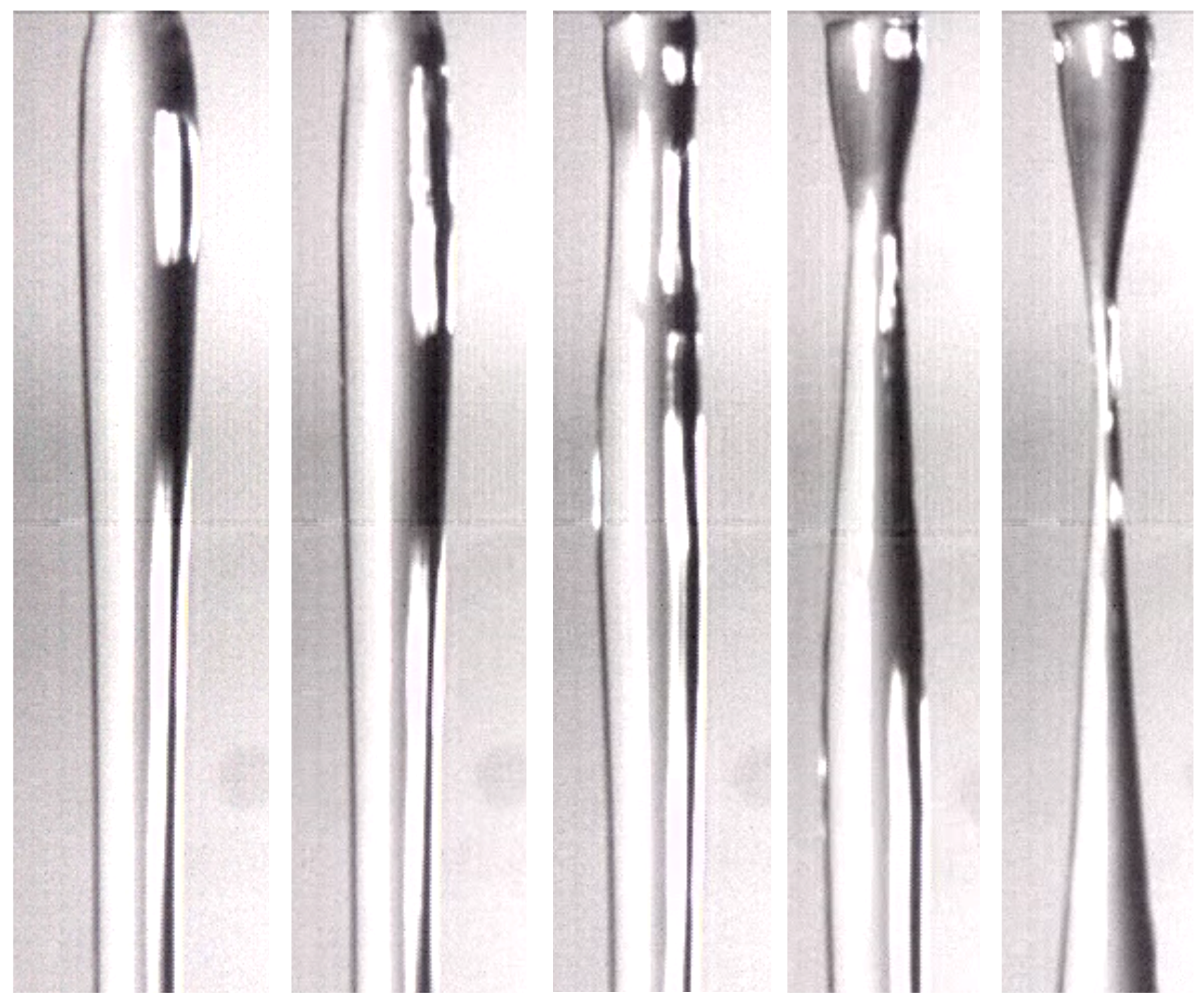}
\caption{}%
\label{rupture}%
\end{center}
\end{figure}

\clearpage
\begin{figure}
[bthp!]
\includegraphics[width=450pt
]
{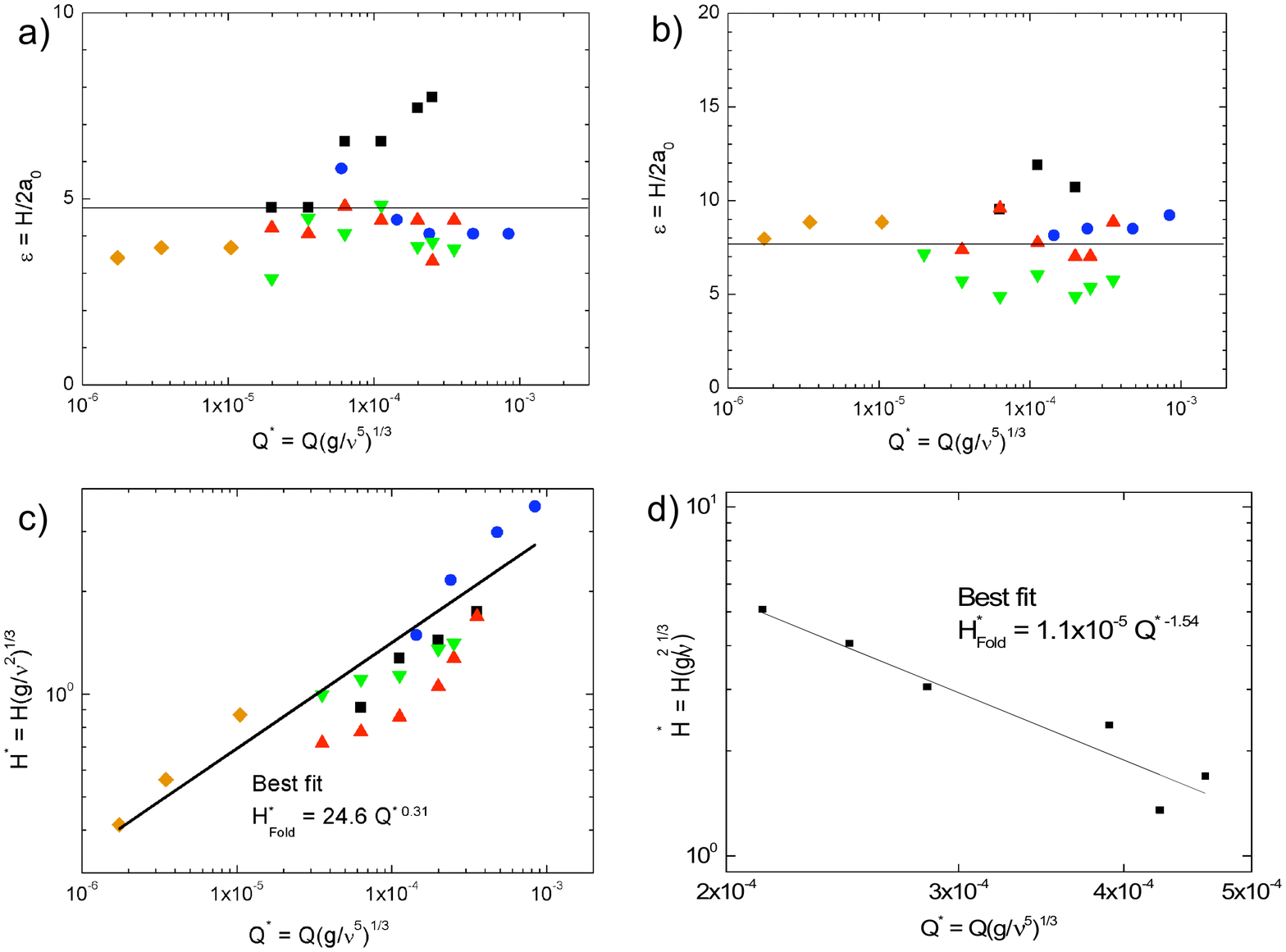}
\caption{}%
\label{allGraphTransitions}%
\end{figure}

\clearpage
\begin{figure}
[bthp!]
\includegraphics[width=450pt]
{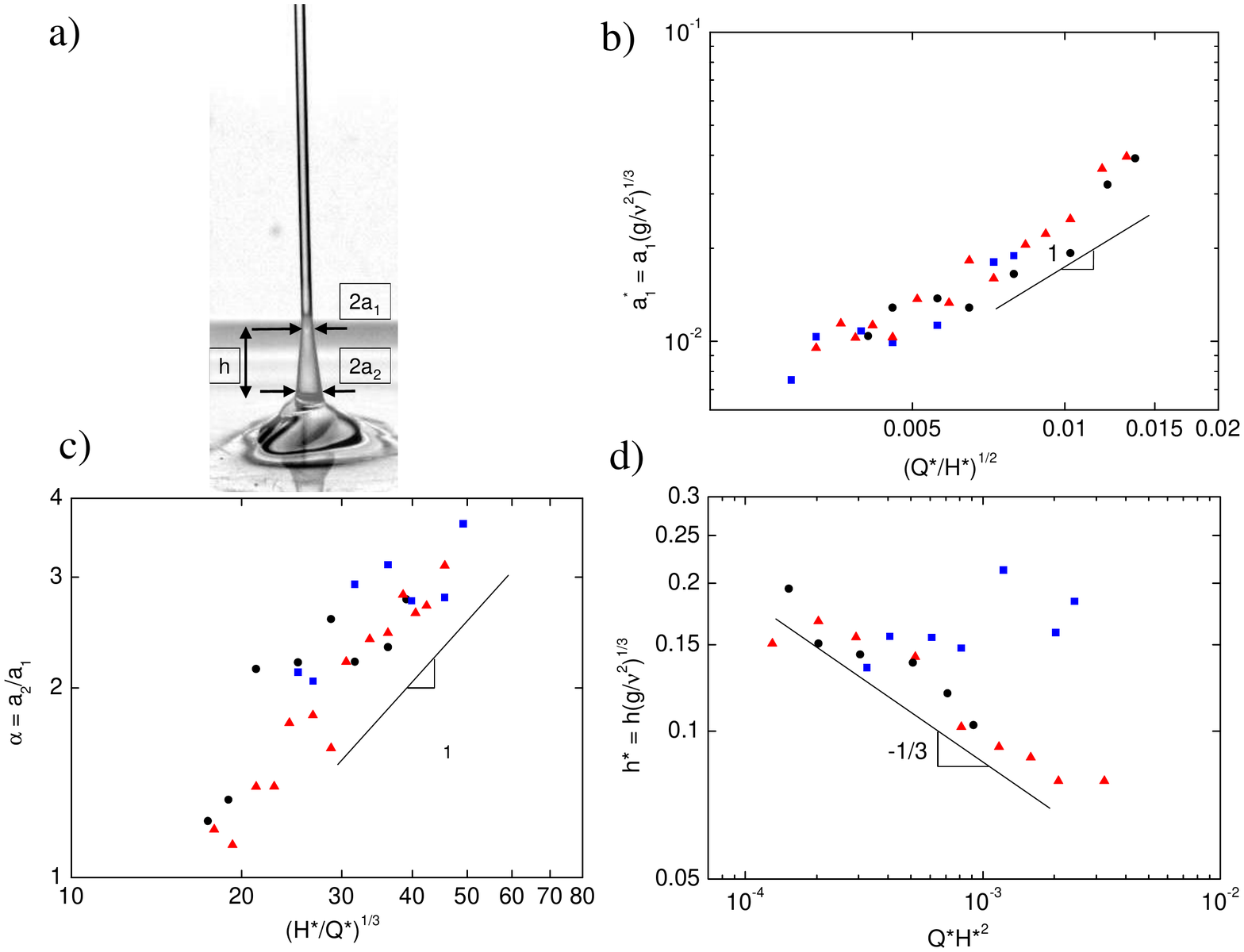}
\caption{}%
\label{allGraphsRadius}%
\end{figure}

\clearpage
\begin{figure}
[bthp!]
\includegraphics[width=450pt]
{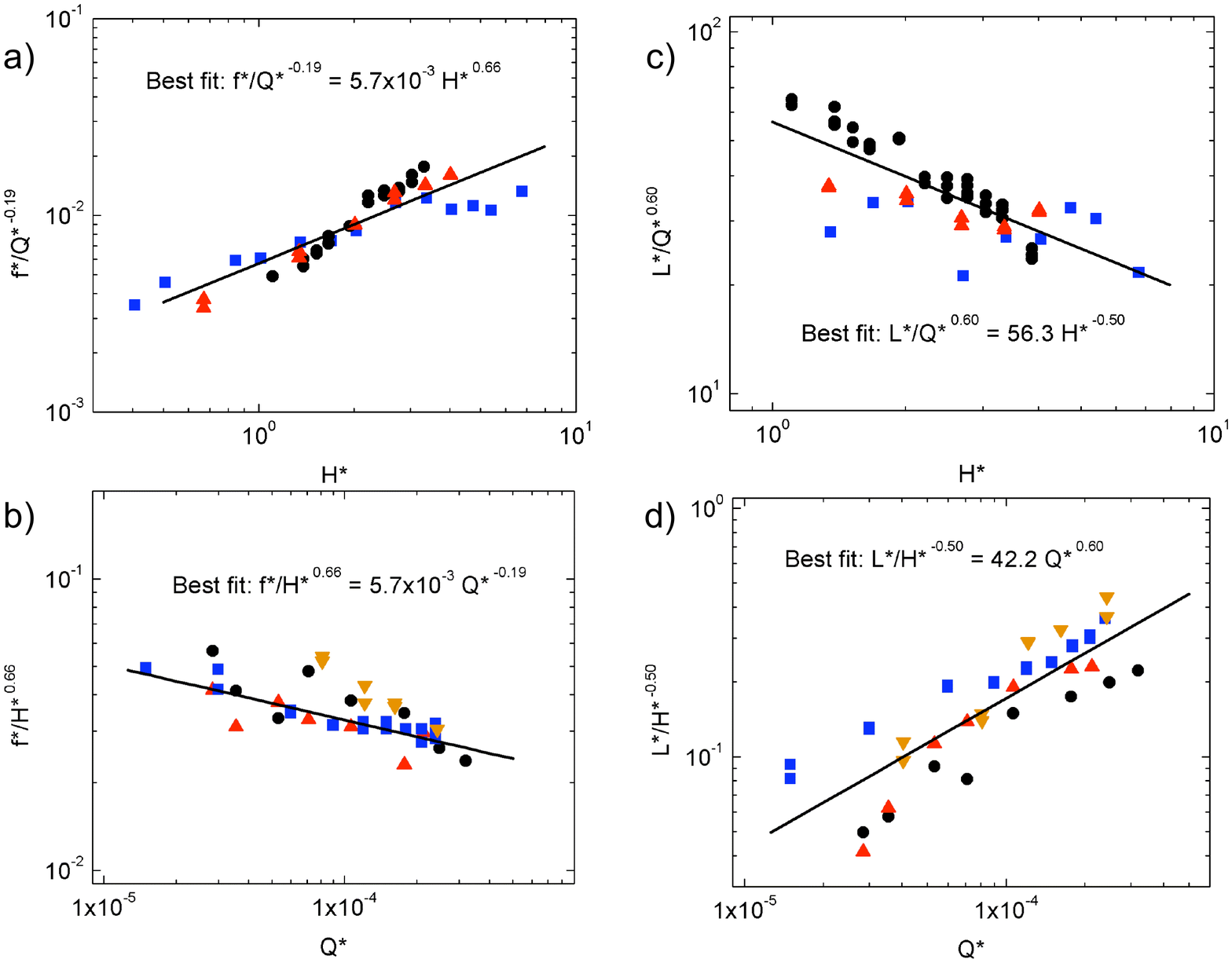}
\caption{}%
\label{allGraphsFrequency}%
\end{figure}


\newpage
\section*{Tables}
\begin{table}[!b]%
\begin{center}
\renewcommand{\tabcolsep}{2pc}\renewcommand{\arraystretch}{1.0}%
\begin{tabular}{|c@{\hspace{12pt}}|l|c@{\hspace{12pt}}|l|c@{\hspace{12pt}}l||}
\hline
{Type of Motion} & {\textbf{Frequency}} & {\textbf{Radius/Amplitude}}\\
\hline
Viscous Coiling & $H^{-1}Qa_{0}^{-2}$ & $H$\\\hline
Gravitational Coiling & $H^{2}Q^{-1/4}$ & $Q^{1/4}$\\\hline
Inertial Coiling & $H^{10/3}Q^{-1/3}$ & $H^{-4/3}Q^{1/3}$\\\hline
Viscous Folding & $H^{-2}Qa_{0}^{-1}$ & $H$\\\hline
Gravitational Folding & $H^{7/6}Q^{-1/6}a_{0}^{-2/3}$ & $H^{-1/2}Q^{1/2}$\\\hline
Inertial Folding & $H^{3/2}Q^{1/6}a_{0}^{-2/3}$ & $H^{-2/3}Q^{1/3}$\\\hline
Experiments Folding & $H^{0.66}Q^{-0.19}$ & $H^{-0.50}Q^{0.60}$\\\hline
\end{tabular}
\end{center}
\caption{Theoretical scaling laws for the different flow regimes,
with respect to the height of the fall, the flow rate and the radius, as well as the experimental results (the last row).}
\label{AllScalingLaws}
\end{table}
\newpage
\begin{table}[!p]%
\begin{center}
\begin{tabular}{|c|c|c|c|c|}
\hline
& \textbf{CPyCl 75} & \textbf{CPyCl 100} & \textbf{CPyCl 150} &
\textbf{Silicone oil}\\\hline
$\lambda$ (s) & 0.61 & 0.72 & 0.90 & 0\\\hline
$\eta_{0}$ (Pa.s) & 8.98 & 18.8 & 54.0 & 9.97\\\hline
$E_{g}$ & 13.3 & 12.4 & 10.9 & 0\\\hline
\end{tabular}
\end{center}

\caption{Viscometric properties of the three CPyCl solutions studied in this work and the silicone oil at
22.5 $^{o}$C.}\label{RheologyData2}%
\end{table}%

\newpage
\begin{table}[!p]%
\begin{center}
\begin{tabular}{|c|c|c|}
\hline
& \textbf{Overall exponent} & \textbf{Range of exponents}\\\hline
$H^{\ast}$ dependence of $f^{\ast}$ & $ 0.66$ & $0.4~<->~1.2$\\
$Q^{\ast}$ dependence of $f^{\ast}$ & $-0.19$ & $-0.18~<->~-0.5$\\\hline
$H^{\ast}$ dependence of $L^{\ast}$&  $-0.50$ & $-0.1~<->~-0.7$\\
$Q^{\ast}$ dependence of $L^{\ast}$& $ 0.60$ & $0.45~<->~0.8$\\\hline
\end{tabular}
\end{center}
\caption{Range of scaling exponents obtained for various series of
experiments, compared to global scaling exponents obtained from regression of full data set.}
\label{datarange}
\end{table}%


\begin{thebibliography}{00}                                                                                        
\bibitem {2008Pouligny} B. Pouligny, M. Chassande-Mottin, Air ingestion by a
buckled viscous jet of silicone oil impacting the free surface of the same
liquid, Phys. Rev. Lett. 100 (2008) 154501.

\bibitem {1982CandMsteady} J. O. Cruickshank, B. R. Munson, The viscous-gravity
jet in stagnation flow, J. Fluid Eng. 104 (1982) 360-362.

\bibitem {1968Taylor}G.I. Taylor, Instability of jets, threads and sheets of
viscous fluid, Proc. 12th Intl. Conf. on Applied Mechanics, Springer, Berlin, 1968. 

\bibitem {1981CandM}J. O. Cruickshank, B. R. Munson, Viscous fluid buckling of
plane and axisymmetric jets, J. Fluid Mech. 113 (1981) 221-239.

\bibitem {1988Cruickshank}J. O. Cruickshank, Low Reynolds number instabilities
in stagnating jet flows, {J. Fluid Mech.} {193} (1988) 111-127.

\bibitem {1993Tchavdarov}B. Tchavdarov, A.L. Yarin and S. Radev, Buckling of
thin liquid jets, {J. Fluid Mech.} {253} (1993) 593-615.

\bibitem {1995Yarin}A. L. Yarin and B. M. Tchavdarov, Onset of folding in
plane liquid films, {J. Fluid Mech.} {307} (1996)  85-99.

\bibitem {1998MahaVJ}L. Mahadevan, W S. Ryu and A D.T. Samuel, Fluid rope
trick investigated, {Nature} {392} (1998) 140-141.

\bibitem {2000MahaCorrection}L. Mahadevan, W.S. Ryu, A. D. T. Samuel, Correction
to ``Fluid rope trick investigated", {Nature} {403} (2000)  502.
 
\bibitem {2004RibeVJ}N. M. Ribe, Coiling of viscous jets, {Proc. R. Soc.
London A} {460} (2004) 3223-3239.

\bibitem {2004RibeVJmultival}M. Maleki, M. Habibi, R. Golestanian, N. M. Ribe, D. Bonn, Liquid rope coiling on a solid surface, {Phys. Rev. Lett.} {93} (2004)  214502.

\bibitem {1988Rehage}H. Rehage and H. Hoffmann, Rheological properties of
viscoelastic surfactant systems, {J. Phys. Chem.} {92} (1988) 4712Ð4719.

\bibitem {1976Israelachvili}J. N. Israelachvili, D. J. Mitchell, B. W. Ninham,
Theory of self-assembly of hydrocarbon amphiphiles into micelles and bilayers,
{J. Chem. Soc. Faraday Trans.} {2} (72) (1976) 1525-1568.

\bibitem {2008Rothstein}J.P. Rothstein, Strong flows of viscoelastic wormlike
micelle solutions, to appear in Rheology Reviews, D.M. Binding and K.Walters
eds., The British Society of Rheology, Aberystwyth, Wales, UK, 2009.

\bibitem {2006McKinley}B. Yesilata, C. Clasen, G.H. McKinley, Nonlinear shear
and extensional flow dynamics of wormlike surfactant solutions, {J.
Non-Newtonian Fluid Mech.} {133} (2006)  73-90.

\bibitem {2003Rothstein}J.P. Rothstein, Transient extensional viscosity of
wormlike micelles solutions,{\ J. Rheology} {47} (2003)  1227-1247.

\bibitem {CPyCl}J.-F. Berret, J. Appell and G. Porte, Linear rheology of
entangled wormlike micelles, {Langmuir} {9} (1993) 2851-2854.

\bibitem {DeGennes}P.-G. de Gennes, Simple views on condensed matter, World Scientific, Singapore, 1992.

\bibitem {Cates}M. E. Cates, Reptation of living polymers: Dynamics of
entangled polymers in the presence of reversible chain-scission reactions,
{Macromolecules} {20} (1987)  2289-2296.

\bibitem {CABERCPyCl}A. Bhardwaj, E. Miller, J.P. Rothstein, Filament
stretching and capillary breakup extensional rheometry measurements of
viscoelastic wormlike micelle solutions, {J. Rheology} {51} (2007) 693-719.

\bibitem {Tanner}R. I. Tanner, Engineering rheology, 2nd ed,  421-435, Oxford
Scientific Press, 2000.

\bibitem {1988Chai}M. S. Chai, Y. L. Yeow, Modeling of fluid M1 using
multiple relaxation time constitutive equations, {J. Non-Newtonian Fluid Mech.} 
{35} (1990)  459-470.

\bibitem{2000MahaVS}M. Skorobogatiy and L. Mahadevan, Folding of viscous sheets and filaments, {Europhys. Lett.} {52} (2000)  532-538.

\bibitem {1993Spenley}N. A. Spenley, M.E. Cates, T.C.B. McLeish, Nonlinear
rheology of wormlike micelles, {Phys. Rev. Lett.} {71} (1993)  6-9.

\bibitem {1994Berret}J.-F. Berret, D. C. Roux, G. Porte, Isotropic-to-nematic
transition in wormlike micelles under shear, {J. Phys. II France} 
{4} (1994) 1261-1279.

\bibitem {DPL}R. B. Bird, R. C. Amstrong, O. Hassager, Dynamics of polymeric
liquids, John Wiley \& Sons, New York, 1987.

\bibitem {CaberMiller}E. Miller, C. Clasen, J. Rothstein, The effect of step-stretch parameters on capillary breakup extensional rheology (CaBER) measurements, {Rheo. Acta} {48} (2009) 625-639.

\bibitem {Larson}R. G. Larson, The structure and rheology of complex fluids,
Oxford University Press, 1999.

\bibitem {IUPACKrafft}PAC, Manual of Symbols and Terminology
for Physicochemical Quantities and Units, Appendix II: Definitions,
Terminology and Symbols in Colloid and Surface Chemistry, {31} (1972) 613.

\bibitem {2006Versluis}M. Versluis, C. Blom, D. van der Meer, K. Van der Weele and
D. Lohse, Leaping shampoo and the stable Kaye effect, {J. Stat. Mech.} {P07} (2006) 007.

\bibitem {2004Lorenceau}E. Lorenceau, J. Eggers, D. Qu\'{e}r\'{e}, Air
Entrainment by a Viscous Jet Plunging into a Bath, {Phys. Rev. Lett.}
{93} (2004)  254501.

\bibitem {BeadsOnString}M. S. N. Oliveira, G. H. McKinley, Iterated stretching
and multiple beads-on-a-string phenomena in dilute solutions of highly
extensible flexible polymers, {Phys. Fluids} {17} (2005)  071704.

\bibitem {ClasenGobbling}C. Clasen, J. Bico, V. Entov, G.H. McKinley,
'Gobbling drops': the jetting/dripping transition in flow of polymer
solutions, {J. Fluid Mech.} {636} (2009)  5-40.

\bibitem {Cromer}M. J. Cromer,  P. L. Cook, G. H. McKinley, Extensional flow of wormlike micellar solutions, {Chem. Eng. Sci.} {64} (2009)  4588-4596.


\bibitem {EntovJNNFM}A. N. Alexandrou, E. Duc, V. Entov, Inertial, viscous, and yield stress effects in Bingham fluid filling of a 2-D cavity, {J. Non-Newtonian Fluid Mech.} { 96} (2001)  383-403.

\end{thebibliography}
\end{document}